\documentclass[fleqn,usenatbib]{mnras}

\usepackage{newtxtext,newtxmath}

\usepackage[T1]{fontenc}
\usepackage{subfigure}
\usepackage{pdflscape}
\usepackage{color}
\usepackage{comment}
\usepackage{graphicx}
\usepackage{amsmath}
\usepackage{amssymb}
\usepackage{booktabs}
\usepackage{makecell}
\usepackage{fancyvrb}
\usepackage{enumerate}

\title[Quiescent galaxy evolution in VANDELS and LEGA-C]{A combined VANDELS and LEGA-C study: the evolution of quiescent galaxy size,
stellar mass and age from $\mathbf{\textit{z} = 0.6}$ to $\mathbf{\textit{z} = 1.3}$}

\author[M. L. Hamadouche et al.]{M. L. Hamadouche$^{1}$\thanks{E-mail: mham@roe.ac.uk (MLH)},
A. C. Carnall$^{1}$,
R. J. McLure$^{1}$, J. S. Dunlop$^{1}$, D. J. McLeod$^{1}$, F. Cullen$^{1}$, R. Begley$^{1}$,
\newauthor M. Bolzonella$^{2}$, F. Buitrago$^{3,4}$, M. Castellano$^{5}$, O. Cucciati$^{2}$, A. Fontana$^{5}$, A. Gargiulo$^{6}$, M. Moresco$^{2,7}$, \newauthor L. Pozzetti$^{2}$ and G. Zamorani$^{2}$ \\
 \\
$^{1}$SUPA\thanks{Scottish Universities Physics Alliance}, Institute for Astronomy, University of Edinburgh, Royal Observatory, Edinburgh, EH9 3HJ, UK\\
$^{2}$INAF - Osservatorio di Astrofisica e Scienza dello Spazio di Bologna, Via Piero Gobetti 93/3, I-40129, Bologna, Italy \\
$^{3}$Departamento de Física Teórica, Atómica y Óptica, Universidad de Valladolid, E-47011 Valladolid, Spain\\
$^{4}$Instituto de Astrofísica e Ciências do Espaço, Universidade de Lisboa, OAL, Tapada da Ajuda, PT1349-018 Lisbon, Portugal\\
$^{5}$INAF - Osservatorio Astronomico di Roma, Via Frascati 33, 00078, Monteporzio Catone, Italy\\
$^{6}$INAF - Istituto di Astrofisica Spaziale e Fisica Cosmica Milano, via A.Corti 12, I-20133 Milano, Italy\\
$^{7}$Dipartimento di Fisica e Astronomia ``Augusto Righi'', Alma Mater Studiorum Universit\`{a} di Bologna, via Piero Gobetti 93/2, I-40129 Bologna, Italy
\vspace{-0.2cm}
}

\date{Accepted XXX. Received YYY; in original form ZZZ}

\pubyear{2021}

\begin{document}
\label{firstpage}
\pagerange{\pageref{firstpage}--\pageref{lastpage}}
\maketitle

\begin{abstract}
We study the relationships between stellar mass, size and age within the quiescent population, using two mass-complete spectroscopic samples with $\mathrm{log_{10}}(M_{\star}/\mathrm{M_{\odot}})>10.3$, taken from VANDELS at $1.0<z<1.3$, and LEGA-C at $0.6<z<0.8$. Using robust D\textsubscript{n}4000 values, we demonstrate that the well-known `downsizing' signature is already in place by $z\simeq1.1$, with D\textsubscript{n}4000 increasing by $\simeq0.1$ across a $\simeq$ 1 dex mass interval for both VANDELS and LEGA-C. We then proceed to investigate the evolution of the quiescent galaxy stellar mass-size relation from $z\simeq1.1$ to $z\simeq0.7$. We find the median size increases by a factor of $1.9\pm{0.1}$ at $\mathrm{log_{10}}(M_{\star}/\mathrm{M_{\odot}})=10.5$, and see tentative evidence for flattening of the relation, finding slopes of $\alpha=0.72\pm0.06$ and $\alpha=$ $0.56\pm0.04$ for VANDELS and LEGA-C respectively. We finally split our sample into galaxies above and below our fitted mass-size relations, to investigate how size and D\textsubscript{n}4000 correlate. For LEGA-C, we see a clear difference, with larger galaxies found to have smaller D\textsubscript{n}4000 at fixed stellar mass. Due to the faintness and smaller numbers of the VANDELS sample, we cannot confirm whether a similar relation exists at $z\simeq1.1$. We consider whether differences in stellar age or metallicity are most likely to drive this size-D\textsubscript{n}4000 relation, finding that any metallicity differences are unlikely to fully explain the observed offset, meaning smaller galaxies must be older than their larger counterparts. We find the observed evolution in size, mass and D\textsubscript{n}4000 across the $\simeq2$ Gyr from $z\sim1.1$ to $z\sim0.7$ can be explained by a simple toy model in which VANDELS galaxies evolve passively, whilst experiencing a series of minor mergers.
\end{abstract}


\section{Introduction}

The formation and quenching of quiescent galaxies is still one of the most debated subjects in extragalactic astronomy. However, the last 20 years has undoubtedly seen major progress, as a result of ever-expanding and increasingly deep photometric and spectroscopic surveys, such as the Sloan Digital Sky Survey \citep[SDSS;][]{yorkSDSS}, zCOSMOS \citep{zcosmos}, the UKIRT Infrared Deep Sky Survey \citep[UKIDSS;][]{ukidss_survey_paper} and the Cosmic Assembly Near-Infrared Deep Extragalactic Legacy Survey \citep[CANDELS;][]{candels_grogin, candels_koekemoer}.

One of the key foundational results that shapes our understanding of galaxy evolution is the galaxy colour bimodality. First quantified in detail using SDSS data in the early 2000s (e.g. \citealt{Strateva2001, Baldry2004}), two distinct sub-populations of galaxies were identified. These are commonly referred to as the ``red sequence'' of quiescent galaxies and ``blue cloud'' of star-forming galaxies. These two populations are bridged by a smaller number of galaxies in transition between the blue cloud and red sequence, widely referred to as ``green valley'' galaxies. The observation of this sharp divide in the local Universe naturally led to questions as to when and why galaxies transition to the red sequence. 

More recently, deep spectroscopic studies have firmly established that quiescent galaxies, as well as the colour bimodality, already exist at least as early as redshift, $z\sim3$ (e.g. \citealt{k20_cimmatti}; \citealt{gdds}; \citealt{gmass_eso}, \citealt{Schreiber2018}). However, photometric studies of the galaxy stellar-mass function (GSMF) have demonstrated that quiescent galaxy number densities have risen rapidly since this time, with quiescent galaxies making up less than 10 per cent of the total mass budget at $z = 3$, rising to around 75 per cent  at $z = 0$ (e.g. \citealt{fontana_downsizing, muzzin_uvista, davidzon_gsmf,mcleod2021_gsmf, santini_2021}).

The existence of the colour bimodality across the majority of cosmic history, coupled with the mass dominance of quiescent galaxies at late times, has firmly established quenching as one of the key questions in galaxy evolution \citep[e.g.][]{PengQuenching, somerville_and_dave_quenching}. However, despite our expanding observational capabilities, it has proven challenging to clearly identify the key processes responsible for quenching, and whether their relative importance changes as a function of redshift.

A wide range of different potential quenching mechanisms have been discussed in the literature. Mechanisms that are directly linked to the internal processes of a galaxy (e.g. quasar-mode and radio-mode AGN feedback) are often characterised as ``internal'' or ``mass'' quenching mechanisms \citep[e.g.][]{crotonAGN, Choi_2018quenchingmech}. In contrast, mechanisms associated with galaxy-galaxy interaction (e.g. mergers, harassment), or interaction between a galaxy and the intracluster medium (e.g. ram pressure stripping) are typically described as ``environmental'' or ``satellite'' quenching. 

To understand how these processes each contribute to galaxy quenching, much effort has been invested in characterising the physical properties of quiescent galaxies across cosmic time, and how these differ from those of star-forming galaxies. One of the most critical results, again derived from SDSS data, was the discovery that local quiescent galaxies follow a steeper relationship between stellar mass and size than local star-forming galaxies, and that, at stellar masses log$_{10}(M_\star/\mathrm{M_\odot}) \leq 11$, quiescent galaxies are smaller than their star-forming counterparts \citep{shen}. 

Motivated by this, significant progress has been made using deep optical and near-infrared imaging surveys to characterise the relationship between the sizes and stellar masses of both quiescent and star-forming galaxies out to high redshifts  \citep[e.g.][]{daddi_2005_passive_evol, trujillo_2006, wu_size_paper, mowla_cosmos_dash,suess_2019a,suess_2019b,size_mass_nedkova}. These studies have demonstrated that different mass-size relations for quiescent and star-forming galaxies were already in place by $z\sim3$. In addition, significant growth is observed in the average sizes of quiescent galaxies over time, increasing by a factor of $\sim2.5$ between $z\simeq1.5$ and $z=0$ \citep[e.g.][]{ross_size_2013, correct_vdw_3dhst+candels}.

It is tempting to interpret increasing average sizes for the quiescent population simply as evidence for the size growth of individual quiescent galaxies, usually assumed to be the result of merger events. However, the situation is complicated by recently quenched star-forming galaxies continuing to arrive on the red sequence over time. As star-forming galaxies are, on average, larger than quiescent galaxies, the addition of these new galaxies to the quiescent population could also plausibly explain this effect \citep[e.g.][]{belli_2015_sizes}. Issues of this nature when comparing similarly selected galaxy samples at different redshifts are commonly referred to as ``progenitor bias'' \citep[e.g.][]{dokkum_franx_prog_bias}. 

Because of the challenges introduced by progenitor bias, a variety of more sophisticated methods have been developed to evolve high-redshift galaxy samples down to the local Universe, with the aspiration of defining evolutionary tracks connecting progenitors and descendants \citep[e.g.][]{zheng_downsizing, van_dokkum_minor_mergers_obs, cimatti_dry_merger, shankar_progenitor_bias, MOSFIREBelli_2019, carnalletal2019, Tacchella2021}. A key parameter that can be introduced to break degeneracies in such analyses is the age of a galaxy's stellar population, or more generally its star-formation history (SFH; e.g. \citealt{Carnall2019b}, \citealt{Leja2019}).

Historically, attempts to constrain the ages of quiescent galaxies have tended to focus on specific spectral features, such as D\textsubscript{n}4000 and the H$\delta$ equivalent width \citep[e.g.][]{orig_dn4000_balogh,kauffman_dn4000}. At \hbox{$z<1$}, many studies have reported a positive correlation between D\textsubscript{n}4000 and stellar mass \citep[e.g.][]{brinchmannMS, zcosmos_moresco10,moresco_cos_chronometers10, moresco_cosmic_hubble_16, haines_d4000, Siudek2017, mass_d4000_kim, Wu2018a}. This has been widely associated with the ``downsizing'' trend, in which the stellar populations of more massive galaxies formed earlier in cosmic history, with present-day star formation concentrated in lower-mass galaxies \citep[e.g.][]{cowie1996downsizing, Thomas2005}.

Recently, the increasing availability of large, representative, spectroscopic samples at intermediate redshifts has facilitated analyses probing how size, age and stellar mass interrelate within the quiescent population. This has been partially motivated by earlier studies, spanning $z = 0 - 2$, that suggest larger galaxies tend to be younger than smaller galaxies within the quiescent population \citep[e.g.][]{correct_vdw_3dhst+candels, belli_2015_sizes,gargiulo_2017}. 

At $z\sim0.7$, \cite{wu_size_paper} report a study of D\textsubscript{n}4000 and H$\delta$ gradients across the stellar mass-size plane for a sample of $\sim$400 H$\beta$-selected quiescent galaxies from the Large Early Galaxy Astrophysics Census (LEGA-C). At fixed mass, a negative correlation is found between D\textsubscript{n}4000 and size, suggesting that larger galaxies are younger than their smaller counterparts. If confirmed, this result provides an opportunity to quantify progenitor bias within the quiescent population, and hence disentangle this effect from the merger-driven size growth of older galaxies post-quenching.

However, it is also well known that D\textsubscript{n}4000 has a significant secondary dependence on stellar metallicity \citep[e.g.][]{bruzualcharlot2003}. Recently, \cite{barone_2018_grav_pot, barone_2021_legac_SAMI} and \cite{beverage_metallicity_not_age} have reported no correlation between age and size at fixed stellar mass, instead suggesting that the observed correlation between D\textsubscript{n}4000 and size is driven by a positive correlation between metallicity and the ratio of stellar mass to radius, acting as a proxy for the depth of a galaxy's gravitational potential well. 

Given this renewed discussion in the recent literature regarding the evolution of the quiescent galaxy population on the stellar mass-size plane, in this work we aim to improve our understanding by exploiting the combined statistical power provided by two recently completed surveys: VANDELS \citep{vandels,pentericcivandels, vandels_final} and LEGA-C \citep[][]{legac_survey_paper, legac_dr3, legac_dr2_release_paper}. These surveys provide ultra-deep spectroscopy for large, representative samples of quiescent galaxies, and we use them to construct mass-complete subsamples with $\mathrm{log_{10}}(M_{\star}/\mathrm{M_{\odot}}) \geq 10.3$, spanning redshift ranges from $0.6~<~z~<~0.8$ and $1.0~<~z~<~1.3$ for LEGA-C and VANDELS, respectively.

We begin by studying the relationship between D\textsubscript{n}4000 and stellar mass via these new large spectroscopic samples. In particular, VANDELS provides the opportunity to confirm whether the clear D\textsubscript{n}4000-mass trend seen at lower redshifts was already in place by $z \gtrsim 1$. We then examine how size correlates with D\textsubscript{n}4000 at fixed mass, and attempt to assess the relative contributions of age and metallicity to any size-D\textsubscript{n}4000 trend. Finally, we discuss the level of progenitor bias and merger activity from $z \sim 1.1$ to $z \sim 0.7$, using a simple toy model to explain the observed evolution.

The structure of this paper is as follows. We first give details of the VANDELS and LEGA-C datasets in Section \ref{uvj}. We then describe our sample selection and fitting methods in Section \ref{sample_selection}. We outline our results describing the correlations between size, D\textsubscript{n}4000 and stellar mass in Section \ref{results}, then discuss these results in Section \ref{discussion}. We summarise our conclusions in Section \ref{conclusions}. All magnitudes are quoted in the AB system, and throughout the paper we use cosmological parameters $H_{0}$ = $\mathrm{70 \ {km} \ {s^{-1}} \ {Mpc^{-1}}}$, $\mathrm{\Omega_{m}}$ = 0.3 and $\mathrm{\Omega_{\Lambda}}$ = 0.7. We assume a \cite{kroupaimf} initial mass function.

\section{DATA}\label{uvj}

\subsection{The VANDELS spectroscopic survey}\label{vandels_data_section}
The VANDELS ESO Public Spectroscopic survey \citep{vandels,pentericcivandels,vandels_final} is an ultra-deep, medium-resolution, optical spectroscopic survey, targeting the Chandra Deep Field South (CDFS) and Ultra Deep Survey (UDS) fields. Data were obtained using the Visible Multi-Object Spectrograph \citep[VIMOS;][]{vimos_le_Fevre} on the ESO Very Large Telescope (VLT). The final VANDELS data release (DR4; \citealt{vandels_final})  provides spectroscopy for $\sim$2100 galaxies in the high-redshift Universe.
VANDELS primarily targeted star-forming galaxies at $z > 2.4$ and massive quiescent galaxies at $1.0\leq z \leq2.5$, with quiescent galaxies making up 13 per cent of the final sample. In total, the survey covers an area of 0.2 deg$^2$. 

\subsubsection{VANDELS Photometric Catalogues and selection criteria}\label{passive_selection_criteria}
The galaxies observed by VANDELS were drawn from four separate photometric catalogues, spanning a UV-NIR wavelength range from \hbox{$0.3{-}5\,\mu$m}. In the central regions of the CDFS and UDS fields, which benefit from CANDELS \textit{HST} imaging, we make use of the catalogues produced by the CANDELS team \citep[][]{galametz_candels_cats,guo_candels_cats}. Two further custom ground-based photometric catalogues were produced by \cite{vandels}, covering the areas immediately surrounding the CANDELS footprints.

The VANDELS parent quiescent sample was selected from these photometric catalogues as follows. Objects were required to have \hbox{$1.0 \leq z\mathrm{_{phot}} \leq 2.5$}, as well as \hbox{\textit{i}-band} magnitudes of $i< 25$, and \hbox{\textit{H}-band} magnitudes of $H \leq 22.5$, corresponding to stellar masses of $\mathrm{log_{10}}(M_{\star}/\mathrm{M_{\odot}}) \gtrsim 10$ \citep[][]{vandels}. Quiescent objects were then selected via additional rest-frame \textit{UVJ} magnitude criteria \citep[e.g.][]{williams09bicolour}. In summary, the VANDELS parent quiescent sample was selected by:

\begin{itemize}
    \setlength\itemsep{0.1em}
    \item $1.0 < z\mathrm{_{phot}} < 2.5$
    \item  $H < 22.5$
    \item $i < 25$
    \item $U - V$ $>$ $0.88 \times (V - J)$ $+$ 0.49 
    \item $U - V > 1.3$
    \item $V - J < 1.6$
    
\end{itemize}
These criteria produce a parent sample of 812 galaxies, of which approximately one third were observed as part of VANDELS.

\subsubsection{VANDELS spectroscopic observations}
In this section, we briefly summarise the VANDELS spectroscopic observations. We refer readers to \cite{pentericcivandels} for the full description. From the parent sample of 812 quiescent galaxies (see Section \ref{passive_selection_criteria}), 281 were randomly assigned slits and observed. Observations were made using the MR grism, which provides a resolution of $R \sim 600$ over a wavelength range from $4800{-}10000$\,\AA. 
All objects were observed for 20, 40 or 80 hours to obtain SNRs of $15{-}20$ per resolution element ($\sim$10\,\AA) in the \textit{i}$-$band. 
Spectroscopic redshifts, $z\mathrm{_{spec}}$, were measured by the VANDELS team, with redshift quality flags assigned as described in \cite{pentericcivandels}. In this paper, we use spectra from the VANDELS DR4 final data release \citep[][]{vandels_final} and consider only those objects with robust spectroscopic redshifts (i.e quality flags 3 and 4, corresponding to a $\simeq 99$ per cent probability of being correct).
This produces an initial sample of 269 galaxies, of which 235 have $z_{\mathrm{spec}} < 1.5$.

\subsection{The LEGA-C spectroscopic survey} \label{legac_data_section}

The Large Early Galaxy Astrophysics Census (LEGA-C) is also an ESO Public Spectroscopy Survey, making use of VIMOS on the VLT. The full survey provides high-quality spectra for a primary sample of $\sim$3000 galaxies between $0.6 \leq z \leq 1.0$, drawn from a $\sim$1.3 deg$^{2}$ area within the UltraVISTA \citep[][]{mccracken2012_uvista} footprint in the COSMOS field. A full description of the survey design and data reduction can be found in \cite{legac_survey_paper} and \cite{legac_dr2_release_paper}.

\subsubsection{LEGA-C photometric catalogues and parent sample} \label{sec:legac_sample_selection}
The LEGA-C primary spectroscopic sample is drawn from a parent photometric sample of $\sim$10,000 galaxies. This parent sample was selected from the UltraVISTA DR1 catalogue of \cite{muzzin_uvista} as follows. Galaxies were first selected to have $0.6 < z < 1.0$, using spectroscopic redshifts where available (primarily from zCOSMOS), or otherwise photometric redshifts measured by \cite{muzzin_uvista}. A redshift-dependent $K_s-$band magnitude limit was then applied, ranging from \hbox{\textit{K}$_{s} < 21.08$} at $z = 0.6$, to \hbox{\textit{K}$_{s} < 20.7$} at $z = 0.8$, to \hbox{\textit{K}$_{s} < 20.36$} at $z = 1.0$. This ensures that galaxies in the LEGA-C sample are sufficiently bright in the observed spectroscopic wavelength range, from $\sim0.6-0.9\,\mu$m.

\subsubsection{LEGA-C spectroscopic observations}
All LEGA-C observations were carried out using the high-resolution (HR-Red) grism, covering a typical wavelength range of $\simeq6000{-}9000$\,{\AA} at a spectral resolution of $R\sim3500$. Each galaxy received $\sim$20 hours of integration time, resulting in an average continuum SNR of $\sim$20\,{\AA}$^{-1}$. The second data release (DR2) consists of 1988 spectra, including 1550 primary targets drawn from the parent sample described in Section \ref{sec:legac_sample_selection}. In this paper, we make use of the LEGA-C DR2 one-dimensional spectra, as well as D\textsubscript{n}4000 measurements, sizes and spectroscopic redshifts. The LEGA-C team prioritise spectroscopic targets by $K_s$-band magnitude, introducing an individual, $K_s$-band-magnitude-dependent sample completeness correction factor, $S_\mathrm{cor}$, for each object. We account for this by calculating LEGA-C median quantities weighted by these $S_\mathrm{cor}$ values throughout our analysis. For full details of the LEGA-C spectroscopic observations, we refer the reader to \cite{legac_dr2_release_paper}.

\begin{table} 
\setlength{\tabcolsep}{6pt}
\renewcommand{\arraystretch}{1.4}
\caption{Details of the parameter ranges and priors adopted for the {\scshape Bagpipes} fitting of both the VANDELS and LEGA-C photometry. Priors listed as logarithmic are uniform in log-base-ten of the parameter.}
\label{tab:bagpipes_model}
    \begin{tabular}{lcccc}
     \midrule
\textbf{Component} & \textbf{Parameter} & \textbf{Symbol / Unit}  & \textbf{Range} & \textbf{Prior}\\ 
  \midrule 
Global & Redshift & $z\mathrm{_{spec}}$ & - & -  \\
\midrule
SFH & \makecell{Stellar mass \\ formed} & $M_{\star}/\mathrm{M_{\odot}}$ & (1,  10$^{13}$) & log  \\
& Metallicity &  $Z_{\star}/\mathrm{Z_{\odot}}$ & 1.0 & -\\
& Falling slope & $\alpha$ & (0.1, 10$^3$) & log   \\
& Rising slope & $\beta$ & (0.1, 10$^3$) & log \\
& Peak time & $\tau $/ Gyr & (0.1,  $t\mathrm{_{obs}}$) & uniform \\ 
\midrule
Dust & \makecell{Attenuation \\ at 5500{\AA} }& $A_{V}$ / mag & (0, 4) & uniform \\
\midrule
\end{tabular}
\end{table}

\begin{figure*}
    \centering
    \includegraphics[width = 0.85\textwidth]{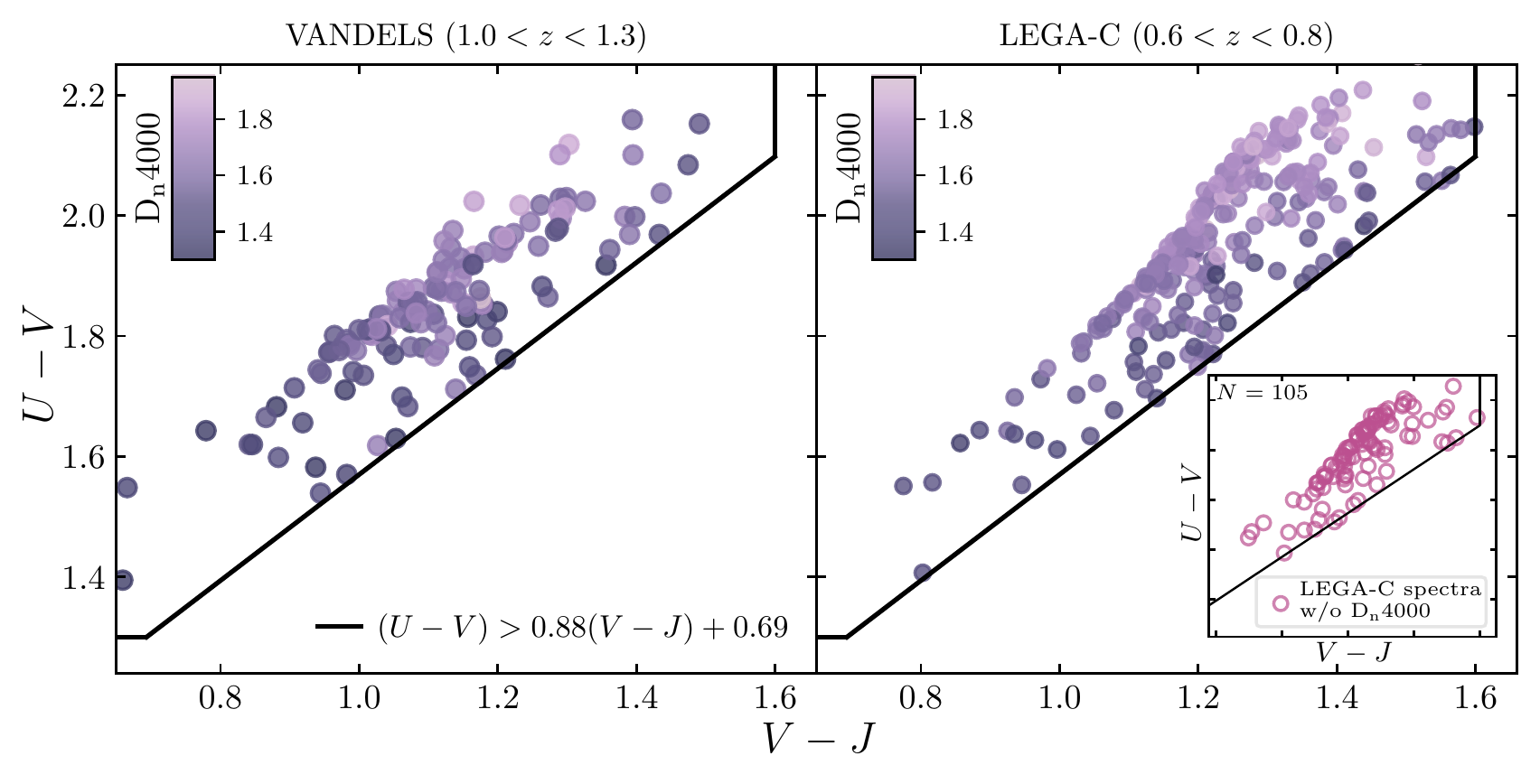}
    \caption{The distribution of our mass-complete (log$_{10}(M_\star/\mathrm{M}_\odot) > 10.3$) quiescent galaxy samples on the \textit{UVJ} plane, using rest-frame colours from {\scshape Bagpipes} \citep{bagpipespaper}, colour-coded by D\textsubscript{n}4000. The VANDELS sample is shown in the left-hand panel, and consists of 137 galaxies (see Section \ref{mass-complete_selection_vandels}). The 272 LEGA-C galaxies with D\textsubscript{n}4000 values are shown in the main part of the right-hand panel, whilst the 105 LEGA-C galaxies without D\textsubscript{n}4000 values are shown in the inset (see Section \ref{legac_mass_complete_sample_selection}). There is a noticeable trend between \textit{UVJ} colour and D\textsubscript{n}4000, with galaxies displaying larger D\textsubscript{n}4000 values at redder colours.}
    \label{fig:legac_vandels_original_uvj}
\end{figure*}

\section{Method and sample selection}\label{sample_selection}

\subsection{Selection of a mass-complete sample from VANDELS}\label{mass-complete_selection_vandels}

To construct a mass-complete sample for our analysis, we begin by fitting the available photometric data for the 269 quiescent galaxies with robust spectroscopic redshifts in VANDELS DR4 (see Section \ref{passive_selection_criteria}), to obtain stellar masses and UVJ colours. We use the {\scshape Bagpipes} code \citep{bagpipespaper}, with the 2016 updated version of the \citealt{bruzualcharlot2003} (BC03) stellar-population-synthesis models \citep[BC16, see][]{charlot_chevallard_2016_BC03}. We adopt the double-power-law SFH model described in \cite{Carnall_2020}, the \cite{actualCalzettiLAW} dust attenuation law, and fixed Solar metallicity for consistency with previous studies (we assume Solar metallicity, $Z_{\odot} = 0.02$). Details of the parameter ranges and priors used are presented in \hbox{Table \ref{tab:bagpipes_model}}.

We first impose stricter UVJ criteria using the {\scshape Bagpipes} colours, requiring \hbox{$U - V > 0.88 \times (V - J) + 0.69$}. This criterion has been shown to consistently select objects with sSFRs less than $0.2\,t_H^{-1}$ \citep{bagpipespaper, carnalletal2019}, where $t_H$ is the age of the Universe, a widely applied criterion for quiescent galaxy selection (e.g. \citealt{pacifici_d4000}). A total of 209 objects meet this UVJ criterion.

The VANDELS quiescent sample is not mass complete across the whole redshift range from $1.0<z<2.5$, and we therefore impose further spectroscopic-redshift and stellar-mass limits to define a mass-complete subsample. Following \cite{carnalletal2019}, we select only objects with redshifts $1.0 \leq z\mathrm{_{spec}} \leq 1.3$ and stellar masses $\mathrm{log_{10}}(M_{\star}/\mathrm{M_{\odot}}) \geq 10.3$ (see their section 3). This narrows our VANDELS sample down to 138 objects. 

In summary, our mass-complete VANDELS sample is selected by:

\begin{itemize}
    \setlength\itemsep{0.5em}
    \item $U - V > 0.88 \times (V - J) + 0.69$
    \item $U - V > 1.3$
    \item $V - J < 1.6$
    \item $1.0 \leq z\mathrm{_{spec}} \leq 1.3$
    \item $\mathrm{log_{10}}(M_{\star}/\mathrm{M_{\odot}}) \geq 10.3$
\end{itemize}

\noindent We finally visually inspect the VANDELS spectra, and remove one further object for which the spectrum is highly contaminated. Our final mass-complete sample therefore consists of 137 galaxies. These are shown on the UVJ diagram in the left-hand panel of Fig. \ref{fig:legac_vandels_original_uvj}.

\subsection{Selection of a mass-complete sample from LEGA-C}\label{legac_mass_complete_sample_selection}

In order to define our final mass-complete LEGA-C quiescent sample, we begin with the 1550 primary objects of the LEGA-C DR2 release (see Section \ref{legac_data_section}). We first exclude objects flagged by the LEGA-C team as having flawed spectra or unreliable redshift measurements. This produced an initial sample of 1212 star-forming and quiescent galaxies between $0.6 < z < 1.0$.

To generate stellar masses and UVJ colours for the LEGA-C galaxies, we use photometry from the updated UltraVISTA DR2 catalogue of \cite{laigle2016}. We cross-match our 1212 LEGA-C galaxies with the \cite{laigle2016} catalogue, finding 1165 matches. We again fit the photometry with {\scshape Bagpipes}, using the same approach described in Section \ref{mass-complete_selection_vandels}. The stellar masses generated by {\scshape Bagpipes} are fully consistent with the masses calculated by the LEGA-C team based on \verb FAST  \citep[][]{kriek2009_fastcode}, as well as the \cite{laigle2016} masses based on \verb LePHARE  \citep{arnouts_lephare, lephare_code_ilbert_06}.

At this point we restricted the redshift range of our LEGA-C sample to $0.6 \leq z \leq 0.8$, meaning that both our VANDELS and LEGA-C samples span similar, $\sim$1 Gyr, periods of cosmic time. This both limits the amount of evolution  taking place within each sample, as well as maximising the time interval between the two samples. LEGA-C is mass complete down to $\mathrm{log_{10}}(M_{\star}/\mathrm{M_{\odot}}) \sim 10$, however we impose a slightly higher mass cut to facilitate direct comparisons with the VANDELS sample, again requiring $\mathrm{log_{10}}(M_{\star}/\mathrm{M_{\odot}}) \geq 10.3$. These two criteria reduce the sample to 656 star-forming and quiescent galaxies. We then apply the same UVJ selection criteria as described in Section \ref{mass-complete_selection_vandels}, resulting in a LEGA-C quiescent sample of 377 galaxies. These are shown in the right-hand panel of Fig. \ref{fig:legac_vandels_original_uvj}.

Due to the truncation of some LEGA-C spectra at the red end, not all LEGA-C galaxies have D\textsubscript{n}4000 measurements in the DR2 catalogue. From our sample of 377 objects, 272 galaxies have measurable D\textsubscript{n}4000 values. One further object has an extremely large uncertainty on D\textsubscript{n}4000, and is therefore excluded from our analysis. In addition, we remove three galaxies for which reliable sizes could not be obtained with \textsc{Galfit} (see Section \ref{size_meas_method}), resulting in a final LEGA-C sample of 268 quiescent galaxies with both size and D\textsubscript{n}4000 information.

\begin{table*}
\centering
\caption{Column one lists the stellar mass range spanned by the three bins employed in Fig. \ref{fig:vandels_mass_stack} and Fig. \ref{fig:legac_mass_stack}. Column two lists the number of VANDELS objects in each bin and column three lists their median stellar mass. Columns four and five list the median D\textsubscript{n}4000 values for objects in each bin, and the D\textsubscript{n}4000 values measured from the stacked spectra generated for objects in each bin, respectively. Columns $9-12$ list the corresponding information for the LEGA-C sample.}
\label{tab:mass_D4000_median_vals}
\setlength{\tabcolsep}{6pt}
\renewcommand{\arraystretch}{1.2}
    \begin{tabular}{ccccccccc}
\toprule
      &  \multicolumn{4}{c}{\textbf{VANDELS}} & \multicolumn{4}{c}{\textbf{LEGA-C}} \\
      \cmidrule(lr){2-5}\cmidrule(lr){6-9}
  Mass Range  & $N$ & $\mathrm{log_{10}}(M_{\star}/\mathrm{M_{\odot}})$ & D\textsubscript{n}4000$\mathrm{_{med}}$ & D\textsubscript{n}4000$\mathrm{_{stack}}$ & $N$ & $\mathrm{log_{10}}(M_{\star}/\mathrm{M_{\odot}})$& D\textsubscript{n}4000$\mathrm{_{med}}$ & D\textsubscript{n}4000$\mathrm{_{stack}}$ \\ \midrule
10.3 $<\mathrm{log_{10}}(M_{\star}/\mathrm{M_{\odot}})  <$ 10.7 & 46& 10.51 $\pm \ 0.02$ &  1.52 $ \pm \ 0.03$ &1.54 $\pm \ 0.01$ & 52 & 10.57 $\pm \ 0.02$&  1.60 $ \pm \ 0.02$ & 1.60 $\pm \ 0.01$ \\
10.7 $< \mathrm{log_{10}}(M_{\star}/\mathrm{M_{\odot}})  <$ 11.1 &69 & 10.90 $\pm \ 0.02$ &  1.59 $ \pm \ 0.02$ & 1.60 $\pm \ 0.01$  & 125 & 10.86 $\pm \ 0.01$  & 1.66 $ \pm \ 0.01$ & 1.64 $\pm \ 0.01$ \\
11.1 $< \mathrm{log_{10}}(M_{\star}/\mathrm{M_{\odot}})  <$ 11.5  & 22& 11.21 $\pm \ 0.03$  & 1.62 $ \pm \ 0.04$ & 1.62 $\pm \ 0.01$& 87& 11.26 $\pm \ 0.01$ & 1.72 $ \pm \ 0.02$ & 1.72 $\pm \ 0.01$  \\
\bottomrule
\end{tabular}
\end{table*}
\begin{figure*}
    \centering
    \includegraphics[width = 0.95\textwidth]{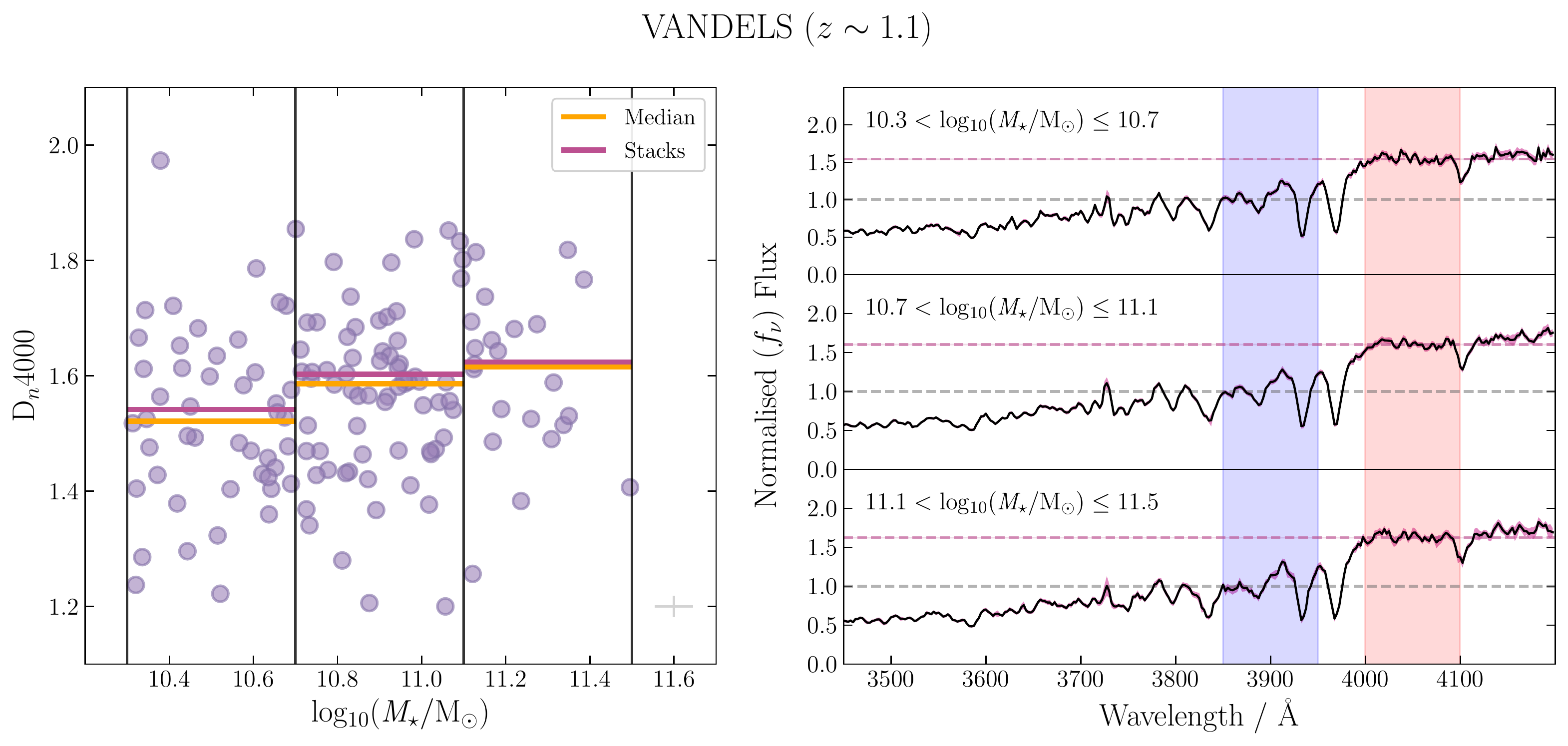}
    \caption{The relationship between stellar mass and D\textsubscript{n}4000 for the VANDELS sample. The left-hand panel shows  D\textsubscript{n}4000 as a function of stellar mass. The right-hand panels show median-stacked spectra in the same three mass bins. The dashed pink lines show the median flux in the red band of the D\textsubscript{n}4000 index. As the spectra have been normalised to the median flux ($f_{\nu}$) in the blue band, the pink lines also corresponds to the D\textsubscript{n}4000 values for each spectrum. We find an increase in D\textsubscript{n}4000 with stellar mass, of $\simeq0.1$ over $\simeq1$ dex in mass, both from the stacked spectra and the median values for individual galaxies (Table \ref{tab:mass_D4000_median_vals}).}
    \label{fig:vandels_mass_stack}
\end{figure*}
\begin{figure*}
    \centering
    \includegraphics[width = 0.95\textwidth]{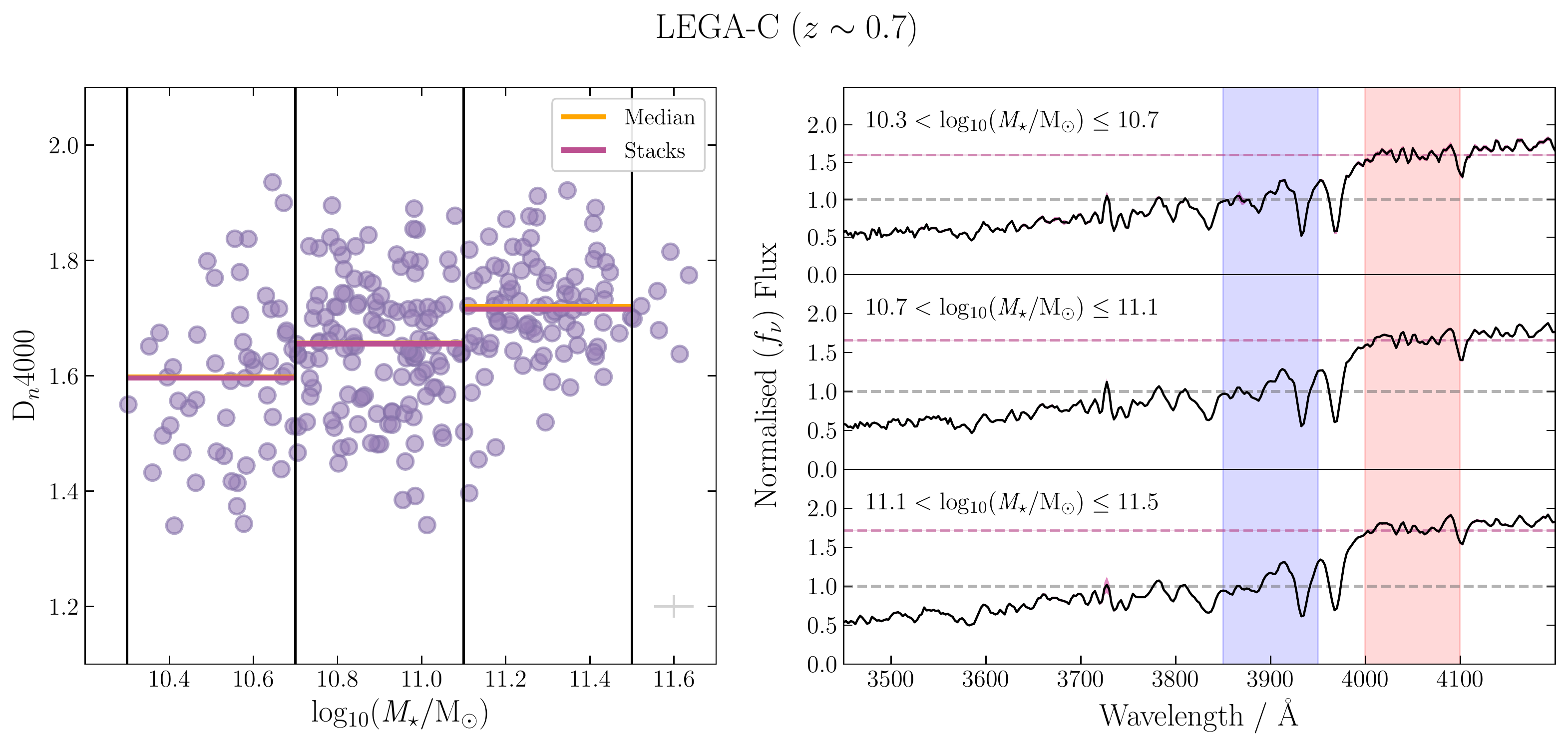}
    \caption{Same as Fig. \ref{fig:vandels_mass_stack} for the LEGA-C sample.}
    \label{fig:legac_mass_stack}
\end{figure*}

\subsection{Measuring galaxy sizes}\label{size_meas_method}

Size measurements for the final sample of VANDELS quiescent galaxies were derived from single S\'ersic fitting of their two-dimensional light profiles using {\sc Galfit}  \citep{galfit_paper}. Within the CANDELS footprint, we use \textit{HST} F160W  imaging for both UDS and CDFS objects. Outside the CANDELS footprints, in CDFS we use \textit{HST} ACS F850LP imaging, and in UDS we use ground-based \textit{H}$-$band imaging from UKIDSS.

For the \textit{HST} imaging, the S\'ersic index (n), effective radius ($r_e$), axis ratio, magnitude and position angle are left as free parameters. For the objects we fitted using the UKIDSS ground-based imaging, we find that the spatial resolution of the \textit{H}$-$band data is not good enough to confidently constrain the S\'ersic index, and we instead assume a constant value of $n=2.5$ during the fitting process\footnote{This value was chosen for consistency with the S\'ersic indices found for galaxies with \textit{HST} imaging. Objects with \textit{HST} F160W imaging have a median S\'ersic index of $n = 2.3$, and a mean of $n=2.7$, suggesting a fixed value of $n=2.5$ for objects with ground-based imaging is a reasonable assumption.}. The rest of the parameters are left free.

We cross-check our size measurements against the results of \cite{vdw_sizes_candels, correct_vdw_3dhst+candels} for common objects, finding good agreement (within $\pm0.1$ dex). In addition, we internally cross-compare the results we obtain for galaxies in each of the four input VANDELS catalogues, demonstrating that the different imaging datasets we employ for size measurement produce consistent results.

For the LEGA-C galaxies, we use the {\sc Galfit} size information provided by the LEGA-C team as part of the DR2 release. These are derived using the methods described in \cite{vdw_sizes_candels} and \cite{legac_survey_paper}, and are based on the original \textit{HST} ACS F814W imaging in COSMOS \citep[][]{scoville}. We note that because the LEGA-C size measurements are based on shorter-wavelength F814W imaging, colour gradients could introduce an offset in sizes relative to those measured for the VANDELS sample. The expected magnitude of this effect is $\simeq 0.05$ dex \citep[e.g.][]{correct_vdw_3dhst+candels}, well within our size-measurement uncertainty. This potentially introduces a small systematic uncertainty into the size evolution calculations presented in Section \ref{discussion}.

\subsection{Measuring \texorpdfstring{D\textsubscript{n}4000}{Measuring Dn4000}}
In this paper, we use the ``narrow'' definition of the 4000\,\AA\ break index, D\textsubscript{n}4000, which has the advantage of being less sensitive to extinction effects \citep[e.g.][]{orig_dn4000_balogh,kauffman_dn4000,martin_2007_d4000,silverman_agn_d4000_zcosmos}. D\textsubscript{n}4000 is calculated as the ratio between the flux per unit frequency in the red continuum band ($4000{-}4100$\,\AA) and the blue continuum band ($3850{-}3950$\,\AA),

\begin{equation}\label{d4000_equation}
    D\mathrm{_{n}}4000 = \dfrac{\langle F^{+} \rangle}{\langle F^{-} \rangle} = \dfrac{(\lambda_{1}^{-} - \lambda_{2}^{-})}{(\lambda_{1}^{+} - \lambda_{2}^{+})}\dfrac{\int_{\lambda_{1}^{+}}^{\lambda_{2}^{+}} f_{\nu} \ d\lambda }{\int_{\lambda_{1}^{-}}^{\lambda_{2}^{-}} f_{\nu} \ d\lambda},
\end{equation}
where $\lambda_{1/2}^{-/+}$ are the upper and lower bounds of the blue ($-$) and red ($+$) narrow continuum bands defined above.

For all galaxies in our final quiescent VANDELS sample, we calculate our own D\textsubscript{n}4000 values directly from the VANDELS spectra, obtaining results for all 137 objects. These have an average uncertainty of 0.02. For the LEGA-C galaxies, we make use of the D\textsubscript{n}4000 values from the LEGA-C DR2 catalogue, which have an average uncertainty of 0.01.

\subsection{Stacked spectra}\label{stacking_analysis}

In order to investigate the average spectral properties of the galaxies in our VANDELS and LEGA-C samples, we construct several median stacked spectra in bins of size and mass. These stacks are all constructed following the same standard procedure. When constructing a stacked spectrum, we first de-redshift and re-sample the individual spectra onto a common wavelength grid, using the {\scshape SpectRes} module for spectral re-sampling \citep{spectres}. We then take the median flux across all the spectra in each pixel, and calculate uncertainties via the robust median absolute deviation (MAD) estimator.

The stacked spectra are normalised by the median flux in the blue continuum band of the 4000\,\AA\,break, in units of erg s$^{-1}$ cm$^{-2}$ Hz$^{-1}$. This means that the median flux density in the blue band lies at \hbox{$f_{\nu}$ = 1.0}, and the median flux in the red continuum band is equal to the value of the D\textsubscript{n}4000 index for that stacked spectrum.

\begin{figure*}
    \includegraphics[width = \textwidth]{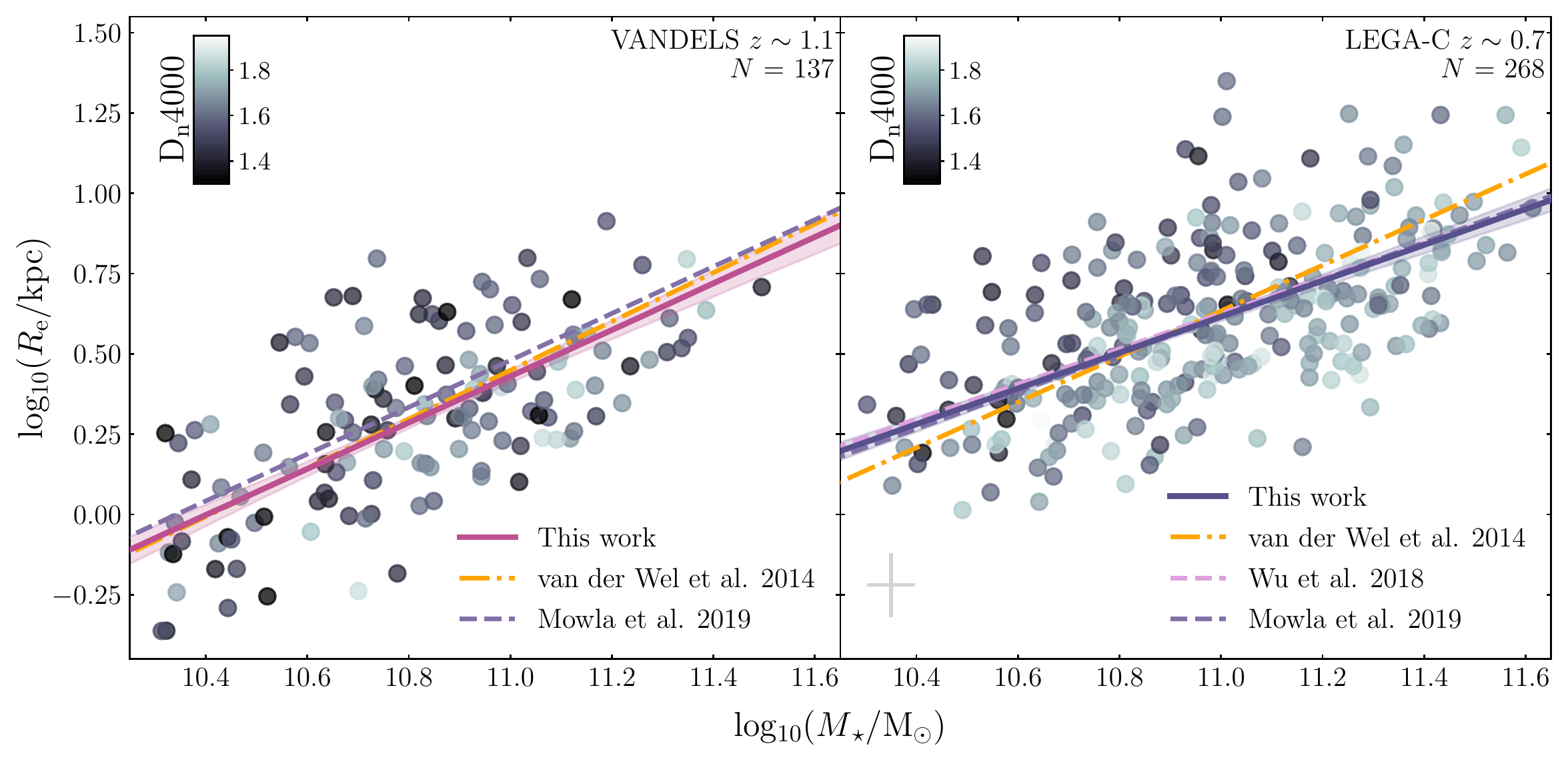}
    \caption{The mass-size distributions of the VANDELS (left) and LEGA-C (right) \textit{UVJ}-selected quiescent samples, colour-coded by D\textsubscript{n}4000. Best-fitting mass-size relations, calculated as described in Section \ref{size-mass-relations}, are shown in each panel. The 1$\sigma$ uncertainties on our best-fitting relations are shown with translucent filled regions. The grey cross shown in the right-hand panel illustrates the typical uncertainties on size and mass for individual objects in both panels. Previous results from the literature for the mass-size relation at both redshifts are shown in both panels.}
    \label{fig:vandels_legac_size_mass}
\end{figure*}

\section{Results}\label{results}

\subsection{The relationship between stellar mass and \texorpdfstring{D\textsubscript{n}4000}{The relationship between stellar mass and Dn4000}}\label{mass-d4000_relation}

In Fig. \ref{fig:vandels_mass_stack}, we show the relationship between stellar mass and D\textsubscript{n}4000 for our mass-complete VANDELS sample (see Section \ref{mass-complete_selection_vandels}). In the left-hand panel, we show D\textsubscript{n}4000 values for each individual galaxy, highlighting the median values in three 0.4-dex stellar-mass bins. In the right-hand panels, we show median stacked spectra in the same three mass bins. The dashed grey and pink lines indicate the median fluxes in the blue and red bands of the D\textsubscript{n}4000 index, respectively. Due to our choice of normalisation, the pink lines correspond to the D\textsubscript{n}4000 values calculated from the stacked spectra.  In Table \ref{tab:mass_D4000_median_vals}, we also present the median D\textsubscript{n}4000 values for galaxies in each mass bin, along with the D\textsubscript{n}4000 values derived from the stacks.

Fig. \ref{fig:legac_mass_stack} shows the relationship between stellar mass and D\textsubscript{n}4000 for our mass-complete LEGA-C sample (see Section \ref{legac_mass_complete_sample_selection}), following the same format as Fig. \ref{fig:vandels_mass_stack}. The median and stacked D\textsubscript{n}4000 values shown in Fig. \ref{fig:legac_mass_stack} for LEGA-C are also listed in Table \ref{tab:mass_D4000_median_vals}.

A clear positive correlation between stellar mass and D\textsubscript{n}4000 is visible in both the VANDELS and LEGA-C samples, with good agreement found in all cases between the median D\textsubscript{n}4000 values for individual objects and the values derived from our stacked spectra. For both VANDELS and LEGA-C, the increase in average D\textsubscript{n}4000 across the mass range we probe is $\simeq0.1$. In addition, similar evolution is observed between VANDELS and LEGA-C at fixed mass in all three bins, with an average evolution in D\textsubscript{n}4000 of $\simeq0.08$ across the $\simeq2$ Gyr that separates our two samples.

\subsection{The VANDELS and LEGA-C mass-size relations} \label{size-mass-relations}

We show our VANDELS and LEGA-C samples on the stellar mass-size plane in Fig. \ref{fig:vandels_legac_size_mass}, together with best-fitting relations of the form:

\begin{equation}
    \mathrm{log}_{10}\bigg(\dfrac{R_e}{\mathrm{kpc}}\bigg) = \alpha \times \mathrm{log}_{10}\bigg(\dfrac{M_\star}{5 \times 10^{10} \mathrm{M}_{\odot}}\bigg) + \mathrm{log}_{10}(A),
    \label{size_mass_eqn}
\end{equation}

\noindent where $R_e$ is the effective radius, $\alpha$ is the slope, and $A$ is the normalisation \citep{correct_vdw_3dhst+candels}. When fitting for the free parameters $(\alpha, A)$, we adopt the stellar-mass uncertainties provided by the {\scshape Bagpipes} spectral fits. However, because the formal errors produced by {\sc Galfit} are known to be significantly underestimated \citep[e.g.][]{haussler_galfit_errs}, we adopt a constant 
error of $\pm 0.1$ dex on the effective radii as a more realistic estimate of the typical uncertainty \citep[e.g.][]{bruce12,ross_size_2013}.
 
For our mass-complete sample of 137 VANDELS quiescent galaxies, we find best-fit parameters of \hbox{$\alpha = 0.72\pm0.06$} and \hbox{log$_{10}(A) = 0.21\pm0.02$}. The best-fitting relationship is shown with a dark pink line in the left-hand panel of Fig. \ref{fig:vandels_legac_size_mass}. The $1\sigma$ confidence interval is shaded pink. Previous $z \sim 1.25$ results from \cite{correct_vdw_3dhst+candels} and \cite{mowla_cosmos_dash} are shown for comparison.

We perform the same fitting on the LEGA-C sample, using only the 272 objects with D\textsubscript{n}4000 values (see Section \ref{legac_mass_complete_sample_selection}). We find a shallower slope than the higher-redshift VANDELS sample, with best-fit parameters of \hbox{$\alpha = 0.56\pm0.04$} and \hbox{log$_{10}(A) =  0.45\pm0.02$}. This relationship is shown with a dark blue line in the right-hand panel of Fig. \ref{fig:vandels_legac_size_mass}. The $1\sigma$ confidence interval is shaded blue. 
Recent literature results at $z \sim 0.75$ from \cite{correct_vdw_3dhst+candels}, \cite{Wu2018a} and \cite{mowla_cosmos_dash} are shown for comparison.

In order to ensure that our results are not biased by adopting the subset of the LEGA-C sample with D\textsubscript{n}4000 measurements, we also fit the full sample of 377 galaxies, obtaining best-fit parameters of \hbox{$\alpha$ = $0.59\pm0.04$} and \hbox{log(A) = $0.42\pm0.02$}. 
The excellent agreement between these results and those quoted above for the D\textsubscript{n}4000 subset means that, in the following 
section, we can compare our VANDELS sample to the LEGA-C subset with D\textsubscript{n}4000 measurements, confident that our results are not sensitive to this choice.

\begin{table*}
\centering
\caption{For both the VANDELS and LEGA-C samples, we present the median values of stellar mass, $\mathrm{log_{10}}(M_{\star}/\mathrm{M_{\odot})}$; size, $\mathrm{log_{10}}(R_{e}/\mathrm{kpc})$, and D\textsubscript{n}4000 within six mass-size bins. The samples are split, first into three 0.4 dex stellar-mass bins, then into galaxies above and below the best-fitting mass-size relations determined in Section \ref{size-mass-relations}. We also list the number of objects within each bin ($N$), and the value of D\textsubscript{n}4000 measured from the corresponding stacked spectrum.}
\label{tab:size_mass_D4000_new}
\setlength{\tabcolsep}{4pt}
\renewcommand{\arraystretch}{1.4}
\begin{tabular}{ccccccccccc}
\toprule
      \textbf{VANDELS} & \multicolumn{5}{c}{\textbf{Galaxies above mass-size relation}} & \multicolumn{5}{c}{\textbf{Galaxies below mass-size relation}} \\
      \cmidrule(lr){1-1}
\cmidrule(lr){2-6}\cmidrule(lr){7-11}
  Mass Range  & $N$  & $\mathrm{log_{10}}\big(\frac{M_{\star}}{\mathrm{M_{\odot}}}\big)$ & D\textsubscript{n}4000$_{\mathrm{med}}$ & D\textsubscript{n}4000$_{\mathrm{stack}}$ & $\mathrm{log_{10}}\big(\frac{R_e}{\mathrm{kpc}}\big)$  & $N$ & $\mathrm{log_{10}}\big(\frac{M_{\star}}{\mathrm{M_{\odot}}}\big)$  & D\textsubscript{n}4000$_{\mathrm{med}}$ & D\textsubscript{n}4000$_{\mathrm{stack}}$ &$\mathrm{log_{10}}\big(\frac{R_e}{\mathrm{kpc}}\big)$ \\ \midrule
10.3 $< \mathrm{log_{10}}\big(\frac{M_{\star}}{\mathrm{M_{\odot}}}\big) \leq$ 10.7 & 22 & 10.56 $\pm \ 0.03$ & 1.56 $\pm \ 0.03$  &1.56 $\pm \ 0.05$  & 0.27 $\pm \ 0.05$ & 24 & 10.46 $\pm \ 0.03$ & 1.49 $\pm \ 0.04$  & 1.50 $\pm \ 0.04$  &-0.07 $\pm \ 0.04$ \\
10.7 $<\mathrm{log_{10}}\big(\frac{M_{\star}}{\mathrm{M_{\odot}}}\big) \leq $ 11.1  & 31 & 10.86 $\pm \ 0.02$ & 1.55 $\pm \ 0.03$  &1.56 $\pm \ 0.04$  & 0.48 $\pm \ 0.03$ & 38 & 10.93 $\pm \ 0.03$ & 1.61 $\pm \ 0.03$  & 1.62 $\pm \ 0.04$ & 0.24 $\pm \ 0.03$ \\
11.1 $< \mathrm{log_{10}}\big(\frac{M_{\star}}{\mathrm{M_{\odot}}}\big) \leq$ 11.5  & 7 & 11.15 $\pm \ 0.04$ & 1.61 $\pm \ 0.08$ &1.64 $\pm \ 0.06$ & 0.67 $\pm \ 0.06$ & 15 & 11.24 $\pm \ 0.03$ & 1.62 $\pm \ 0.04$  &1.60 $\pm \ 0.04$  & 0.48 $\pm \ 0.04$ \\
\bottomrule
\\
\toprule
       \textbf{LEGA-C} & \multicolumn{5}{c}{\textbf{Galaxies above mass-size relation}} & \multicolumn{5}{c}{\textbf{Galaxies below mass-size relation}} \\
       \cmidrule(lr){1-1}\cmidrule(lr){2-6}\cmidrule(lr){7-11}
Mass Range & $N$ & $\mathrm{log_{10}}\big(\frac{M_{\star}}{\mathrm{M_{\odot}}}\big)$ & D\textsubscript{n}4000$_{\mathrm{med}}$ & D\textsubscript{n}4000$_{\mathrm{stack}}$ &$\mathrm{log_{10}}\big(\frac{R_e}{\mathrm{kpc}}\big)$  & $N$ & $\mathrm{log_{10}}\big(\frac{M_{\star}}{\mathrm{M_{\odot}}}\big)$  & D\textsubscript{n}4000$_{\mathrm{med}}$ & D\textsubscript{n}4000$_{\mathrm{stack}}$ &  $\mathrm{log_{10}}\big(\frac{R_e}{\mathrm{kpc}}\big)$ \\ \midrule
10.3 $< \mathrm{log_{10}}\big(\frac{M_{\star}}{\mathrm{M_{\odot}}}\big)\leq$ 10.7 & 24 & 10.51 $\pm \ 0.03$ & 1.55 $\pm \ 0.03$ &1.53 $\pm \ 0.03$ & 0.50 $\pm \ 0.04$ & 27 & 10.56 $\pm \ 0.02$ & 1.65 $\pm \ 0.02$   &1.64 $\pm \ 0.03$   &0.21 $\pm \ 0.02$\\
10.7 $< \mathrm{log_{10}}\big(\frac{M_{\star}}{\mathrm{M_{\odot}}}\big) \leq $ 11.1  & 64 & 10.85 $\pm \ 0.02$ & 1.62 $\pm \ 0.02$  &1.59 $\pm \ 0.02$  &0.68 $\pm \ 0.03$& 60 & 10.87 $\pm \ 0.02$ & 1.70 $\pm \ 0.02$  &1.68 $\pm \ 0.02$  & 0.40$\pm \ 0.02$\\
11.1 $< \mathrm{log_{10}}\big(\frac{M_{\star}}{\mathrm{M_{\odot}}}\big) \leq$ 11.5  & 33 & 11.28 $\pm \ 0.03$ & 1.68 $\pm \ 0.01$  & 1.67 $\pm \ 0.02$  & 0.91 $\pm \ 0.03$ & 54 & 11.25 $\pm \ 0.02$ & 1.73 $\pm \ 0.02$  & 1.71 $\pm \ 0.01$  & 0.65 $\pm \ 0.01$ \\
\bottomrule
\end{tabular}
\end{table*}

\begin{figure*}
\centering
\includegraphics[width=\textwidth]{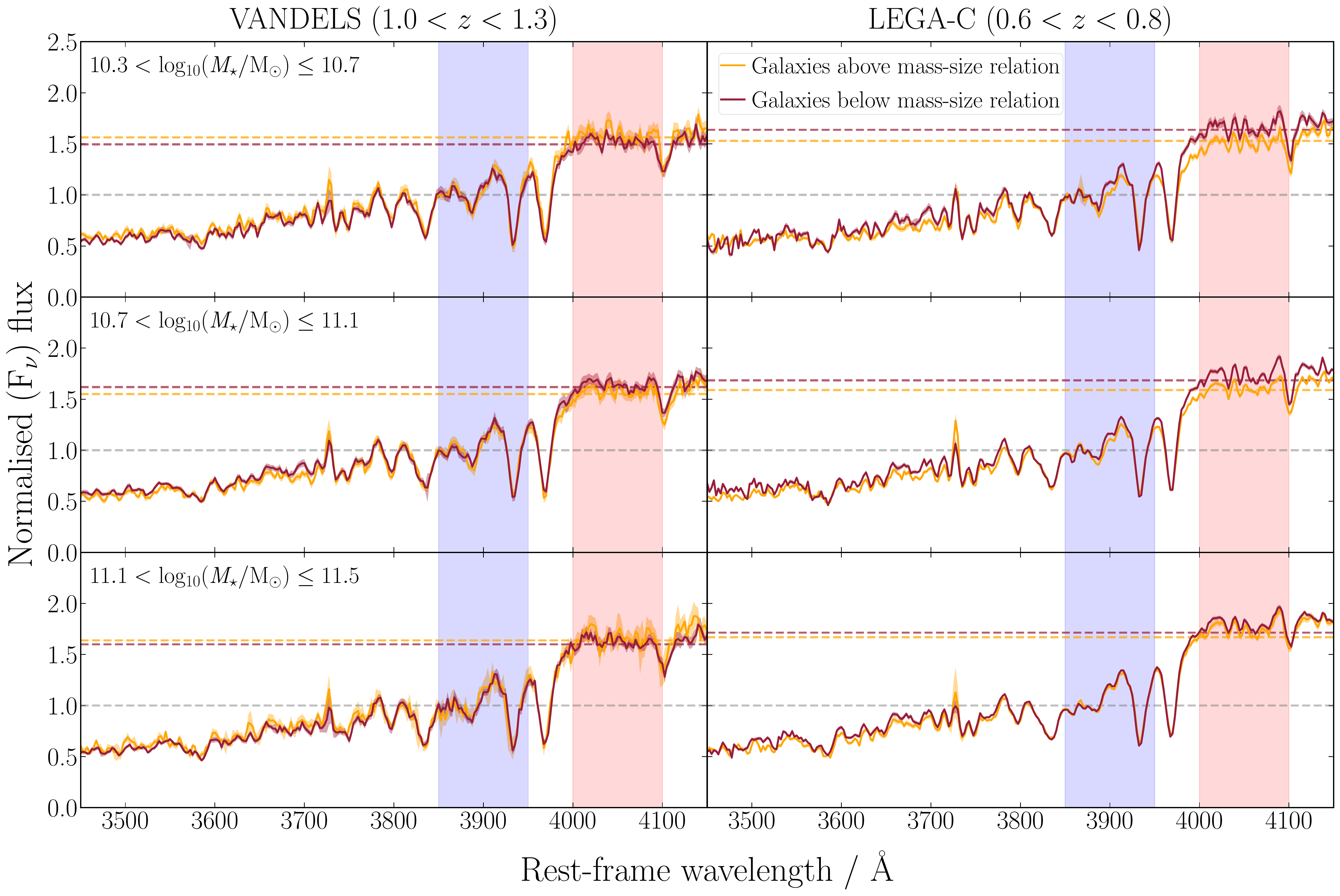}
\caption{The stacked spectra for galaxies above and below the best-fitting mass-size relations for the VANDELS (left-hand panels) and LEGA-C (right-hand panels) samples, within three 0.4-dex stellar-mass bins.
The stacked spectra are created using the methods described in Section \ref{stacking_analysis}.
In all panels, the dashed red line corresponds to the D\textsubscript{n}4000 value calculated from the stacked spectrum below the mass-size relation, and the dashed orange line corresponds to the D\textsubscript{n}4000 value calculated from the stacked spectrum above the mass-size relation. The uncertainties on the measured D\textsubscript{n}4000 values are indicated with coloured shading.}
\label{fig:new_size_mass_stack}
\end{figure*}

\subsection{Stellar mass-size trends with \texorpdfstring{D\textsubscript{n}4000}{Dn4000}}\label{size_dn4000_results}

Having considered the mass-D\textsubscript{n}4000 and mass-size relations for our samples separately in Sections \ref{mass-d4000_relation} and \ref{size-mass-relations}, we now combine these analyses to consider the relationship between galaxy size and D\textsubscript{n}4000. 
Given that both quantities depend strongly on stellar mass, it is critical to separate out this dependency as much as possible (e.g. by using narrow stellar-mass bins) in order to delineate any correlation between D\textsubscript{n}4000 and size.

The VANDELS and LEGA-C samples are shown on the mass-size plane in Fig. \ref{fig:vandels_legac_size_mass}, colour-coded by D\textsubscript{n}4000. 
For the LEGA-C sample at $z\sim0.7$, a vertical trend appears to be visible, with galaxies below the mass-size relation tending to have higher D\textsubscript{n}4000 values (lighter colours) than galaxies above the relation. 
For the VANDELS sample, a corresponding trend is not clearly visible. This is perhaps not surprising, given the smaller number of VANDELS objects, and the larger uncertainties on the individual D\textsubscript{n}4000 measurements due to the lower SNR of the VANDELS spectra.

Motivated by the apparent size-D\textsubscript{n}4000 trend visible for \hbox{LEGA-C} in Fig. \ref{fig:vandels_legac_size_mass}, we implement an approach similar to that of \hbox{Section \ref{mass-d4000_relation}}, splitting our VANDELS and LEGA-C samples into the same three 0.4-dex stellar-mass bins, before splitting each mass bin into two size bins. Galaxies are separated in size by their position relative to the best-fitting mass-size relations derived in Section \ref{size-mass-relations}, with those above being separated from those below. This results in three mass bins for galaxies above the mass-size relation, and three mass bins for galaxies below the mass-size relation.

As in \hbox{Section \ref{mass-d4000_relation}}, we generate stacked spectra for each mass-size bin, from which we measure D\textsubscript{n}4000 values. The six stacked spectra generated for our VANDELS and LEGA-C samples are shown in Fig. \ref{fig:new_size_mass_stack}. The D\textsubscript{n}4000 values for each stack are listed in \hbox{Table \ref{tab:size_mass_D4000_new}}, along with median D\textsubscript{n}4000 values for the individual galaxies entering each stack. From Fig. \ref{fig:new_size_mass_stack} and Table \ref{tab:size_mass_D4000_new}, it can be seen that LEGA-C galaxies lying below the mass-size relation have significantly higher D\textsubscript{n}4000 values than those above the relation. Smaller-than-average galaxies have higher D\textsubscript{n}4000 by $\simeq0.1$ in the lower two mass bins, whereas the difference is $\simeq0.05$ in the highest-mass bin. This offset is most significant ($\simeq4\sigma$) in the middle-mass bin, where the number of objects is greatest, but is also clearly visible ($\simeq2\sigma$ significance) in 
the highest-mass and lowest-mass bins.

For VANDELS, we do not observe a significant difference in average D\textsubscript{n}4000 values above and below the mass-size relation in any of our three stellar-mass bins. This could of course simply indicate that the size-D\textsubscript{n}4000 trend observed in LEGA-C was not in place at $z\sim1.1$. However, as discussed above, the smaller number of objects in the VANDELS sample, coupled with the individual spectra having lower SNRs, makes it difficult to draw any firm conclusions. For example, it can be seen from the uncertainties quoted in Table \ref{tab:size_mass_D4000_new} that, in a given stellar-mass bin, the difference in D\textsubscript{n}4000 above and below the mass-size relation would have to be $\geq 0.15$ in order to be detected with $\geq 2\sigma$ significance in the VANDELS sample.

\section{Discussion} \label{discussion}

In Section \ref{results}, we have considered the relationships between stellar mass, size and D\textsubscript{n}4000 for two mass-complete samples of quiescent galaxies with $\mathrm{log_{10}}(M_{\star}/\mathrm{M_{\odot})} > 10.3$, at $0.6 < z < 0.8$ from \hbox{LEGA-C}, and at $1.0 < z < 1.3$ from VANDELS. In this section we discuss these results, with a particular focus on whether the observed trends in D\textsubscript{n}4000 are driven by stellar age or metallicity, and also on understanding the changes in the quiescent population across the \hbox{$\simeq2$ Gyr} that separates the VANDELS and LEGA-C samples.

\subsection{\texorpdfstring{Evidence of downsizing from stellar mass vs D\textsubscript{n}4000}{Evidence of downsizing from stellar mass vs Dn4000}}\label{stellar_mass_d4000_trends_discussion}

In Section \ref{mass-d4000_relation}, we report a clear relationship between stellar mass and D\textsubscript{n}4000 for both our VANDELS and LEGA-C samples, such that, on average, more massive galaxies have higher D\textsubscript{n}4000. This is in good agreement with previous studies at $z < 1$ \citep[e.g.][]{kauffman_dn4000, brinchmannMS, haines_d4000, mass_d4000_kim, Wu2018a}, and has been widely interpreted as evidence that more massive galaxies are older (often referred to as downsizing, or archaeological downsizing). Our VANDELS result demonstrates this downsizing signal was already in place at $z\gtrsim1$ for the first time using a large, representative, spectroscopic sample. 

In this context, the trends between D\textsubscript{n}4000 and stellar mass seen in Fig. \ref{fig:vandels_mass_stack} and Fig. \ref{fig:legac_mass_stack} can be securely interpreted in terms of stellar populations of different ages, given the broad agreement in the literature that the stellar mass vs stellar metallicity relation is relatively flat at $\mathrm{log_{10}}(M_{\star}/\mathrm{M_{\odot})} > 10$ (e.g. \citealt{Gallazzi2005, Gallazzi2014, Panter2008, choi_2014_mzr, leethochawalit_z_0.5_mzr, adam_metallicity_21}). We also see an increase in D\textsubscript{n}4000 between VANDELS and LEGA-C, of $\simeq0.06{-}0.10$ at fixed stellar mass. This redshift evolution is discussed in Section \ref{toy_model_section}.

\subsection{Stellar mass-size relations}\label{size-mass-discussion}

The stellar mass-size relations followed by both star-forming and quiescent galaxies have been extensively investigated over the past two decades. Over the widest dynamic range in stellar mass, there is now good evidence that quiescent galaxies follow a mass-size relation that is well described by a double power law \citep[e.g.][]{size_mass_nedkova,subaru_cam_size_mass}. However, at high stellar masses (i.e. $\mathrm{log_{10}}(M_{\star}\mathrm{/M_{\odot})} > 10$), it is clear that quiescent galaxies follow a linear $\log_{10}({R_{e}})-\log_{10}({M_{\star}})$ relation (see Equation \ref{size_mass_eqn}), with a normalisation that evolves rapidly with redshift \citep[e.g.][]{shen,buitrago_2008, ross_size_2013, correct_vdw_3dhst+candels, wu_size_paper, mowla_cosmos_dash}. It is much less clear whether the slope of the mass-size relation also evolves with time: most studies report values in the range \hbox{$0.50<\alpha<0.75$}, with no clear redshift trend emerging.

The mass-size relations we derive in this work, based on robust, mass-complete, spectroscopically confirmed galaxy samples at \hbox{$z\sim 0.7$} and  \hbox{$z\sim 1.1$}, are consistent with recent literature results based on larger galaxy samples based on photometric redshifts. Our results indicate that, at a given stellar mass, massive quiescent galaxies grow by a factor of $\simeq 2$ over the $\simeq 2$ Gyr of cosmic time that separates our two samples. 

Our results also provide tentative evidence that the slope of the mass-size relation may flatten ($\Delta \alpha =-0.16\pm{0.07}$) over the same interval, although we note that any flattening is only significant at the $\simeq 2\sigma$ level. 
This tentative flattening of the relation could potentially be explained by larger newly-quenched galaxies, arriving onto the stellar mass-size distribution between $10.3~\leq~ \mathrm{log_{10}}(M_{\star}/\mathrm{M_{\odot}})~\leq ~10.65$.
In Section \ref{toy_model_section}, we explore whether a simple minor-merger model can explain the evolution of the quiescent galaxy mass-size relation suggested by our VANDELS and LEGA-C galaxy samples.

\begin{figure}
    \centering
    \includegraphics[width=\columnwidth]{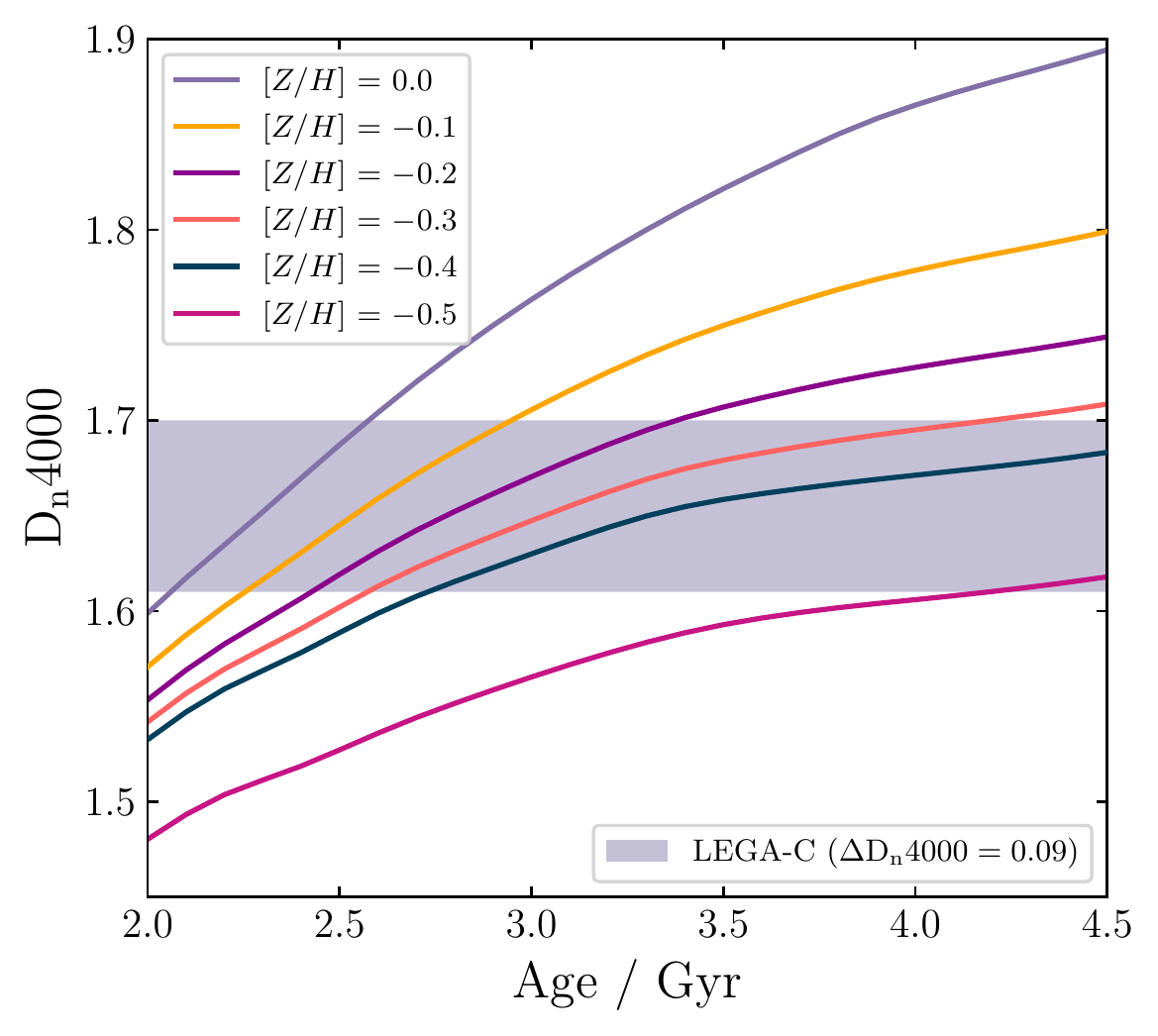}
    \caption{The relationship between D\textsubscript{n}4000, mean stellar age and stellar metallicity. We illustrate the way that D\textsubscript{n}4000 increases with mean stellar age for six fixed metallicities, assuming 1-Gyr-duration constant star-formation histories. The shaded region shows the difference between the D\textsubscript{n}4000 values for the two size bins within the middle mass bin of the LEGA-C sample. Within the range of plausible metallicities (i.e. $0.0 <$ [Z/H] $< -0.3$), it can be seen that an age difference of $\simeq0.5{-}1$ Gyr, or a metallicity difference of $\simeq0.3{-}0.4$ dex would be required to explain the observed D\textsubscript{n}4000 offset. Note that we assume a value for Solar metallicity of $Z_\odot=0.02$.}
    \label{fig:ssp_age_metallicity}
\end{figure}

\subsection{\texorpdfstring{The size-D\textsubscript{n}4000 relation: changing age or metallicity?}{Dn4000}}\label{ssp_metallicity_age}

The spectral index D\textsubscript{n}4000 is best known as a way of inferring ages for older stellar populations (e.g. \citealt{orig_dn4000_balogh,kauffman_dn4000}). However, the strength of this break is also known to depend somewhat on metallicity \citep[e.g.][]{bruzualcharlot2003}.

In \cite{wu_size_paper}, the authors report a size-D\textsubscript{n}4000 relationship for a similar sample of LEGA-C quiescent galaxies, and suggest that this is driven by a difference in the average ages of larger and smaller objects. However, other studies \citep[e.g.][]{barone_2021_legac_SAMI, beverage_metallicity_not_age} have reported a size-metallicity correlation via full spectral fitting analyses, casting doubt as to whether any size-D\textsubscript{n}4000 trend should be attributed primarily to differences in average age or metallicity.

In Section \ref{size_dn4000_results}, we have established that there is a trend between size and D\textsubscript{n}4000 for our LEGA-C sample, with smaller galaxies having higher D\textsubscript{n}4000. Averaged over our three mass bins, the mean D\textsubscript{n}4000 offset is $\simeq0.08$, in good agreement with \cite{wu_size_paper}. 

To investigate the influence of age and metallicity on D\textsubscript{n}4000 in the context of our LEGA-C sample, we use {\scshape Bagpipes} to run a grid of models with simple, 1-Gyr-duration constant star-formation histories, using the BC03 stellar-population models. In Fig. \ref{fig:ssp_age_metallicity}, we show the relationship between D\textsubscript{n}4000 and mean stellar age, for a range of metallicities (note that we assume $Z_\odot = 0.02$). Also shown with indigo shading is the gap between the median D\textsubscript{n}4000 values above and below the mass-size relation in our middle LEGA-C stellar-mass bin. This mass bin was chosen as it contains the most objects, and hence has the best-constrained median D\textsubscript{n}4000 values.

Under this simple set of assumptions, and within the range of plausible metallicities (i.e. $0.0 <$ [Z/H] $< -0.3$), it can be seen that our $\Delta$D\textsubscript{n}4000 of 0.09 suggests a difference in age of $\simeq0.5{-}1$ Gyr, if the average metallicities in the two size bins are the same (the horizontal distance over which the lines shown move from the bottom to the top of the shaded region). Conversely, if both bins contain objects with the same average ages, a metallicity offset of $\sim0.3{-}0.4$ dex is required to explain the observed difference in D\textsubscript{n}4000 values.

Several leading studies have reported that massive, quiescent galaxies at $z\sim0.7$ exhibit a scatter in metallicity of $\simeq0.2$ dex (\citealt{Gallazzi2014}, \citealt{Tacchella2021}). Even in the most extreme possible scenario, for which all galaxies above the mass-size relation have below-average metallicities and vice versa, this level of scatter is not sufficient to produce the $0.3{-}0.4$ dex metallicity offset required to fully explain the $\Delta$D\textsubscript{n}4000 = 0.09 observed in our middle LEGA-C mass bin. We therefore conclude that the size-metallicity relations found by \cite{barone_2021_legac_SAMI} and \cite{beverage_metallicity_not_age} cannot fully explain our observed size-D\textsubscript{n}4000 trend, meaning a size-age relation is likely to also be present in our LEGA-C sample, with a magnitude of $\lesssim$500 Myr.

As discussed in Section \ref{size_dn4000_results}, we do not observe a significant size-D\textsubscript{n}4000 relation for our VANDELS sample. From Fig. \ref{fig:ssp_age_metallicity}, it can be seen that any size-D\textsubscript{n}4000 trend due to a fixed offset in age or metallicity would be smaller in a population of younger galaxies. We therefore would not expect to see clear evidence of a size-age or size-metallicity relation in VANDELS with similar magnitude to the one we observe for LEGA-C, given the larger uncertainties on our binned VANDELS median D\textsubscript{n}4000 values.

\subsection{A model linking the VANDELS and LEGA-C samples}\label{toy_model_section}

Having established the level of evolution between the LEGA-C and VANDELS samples in terms of stellar mass, size and D\textsubscript{n}4000, in this section we investigate whether or not the observed evolution is compatible with the simplest possible toy model. Specifically, we investigate whether the observed size and D\textsubscript{n}4000 evolution can be explained by passively evolving the VANDELS galaxies by $\simeq 2$ Gyr at constant metallicity, with the growth in stellar mass and size being the result of a series of minor mergers.

\subsubsection{Accounting for progenitor bias}\label{toy_model_number_densities}
In order to gain an accurate understanding of the relationship between the VANDELS and LEGA-C datasets, we first calculate the fraction of the LEGA-C sample that are likely descendants of the higher-redshift VANDELS galaxies, thus constraining the level of progenitor bias. Based on the original parent samples for both \hbox{LEGA-C} and VANDELS, we estimate that the comoving number densities of quiescent galaxies with $\mathrm{log_{10}}(M_{\star}/\mathrm{M_{\odot})} \geq 10.3$ at $z\sim0.7$ and $z\sim1.1$ are $9.10 \times 10^{-4} \ \mathrm{Mpc}^{-3}$ and $7.62 \times 10^{-4} \ \mathrm{Mpc}^{-3}$, respectively. As a result, we conclude that the VANDELS sample can account for $\simeq 85$ per cent of the progenitors of the LEGA-C sample. 

Alternatively, we can map the VANDELS comoving number density to a higher-mass subset of the LEGA-C sample. We make the simplifying assumption here that the most massive objects in LEGA-C have been quenched long enough to be descended from the VANDELS sample at $z \sim 1.1$. Under this assumption, the VANDELS sample can account for 100 per cent of the progenitors of the LEGA-C sample with stellar masses $\mathrm{log_{10}}(M_{\star}/\mathrm{M_{\odot})} \geq 10.65$. We adopt this latter assumption throughout the rest of this section, comparing the whole VANDELS sample ($\mathrm{log_{10}}(M_{\star}/\mathrm{M_{\odot})} \geq 10.3$) with the LEGA-C sample at $\mathrm{log_{10}}(M_{\star}/\mathrm{M_{\odot})} \geq 10.65$.

\begin{figure}
   \centering
    \includegraphics[width = \columnwidth]{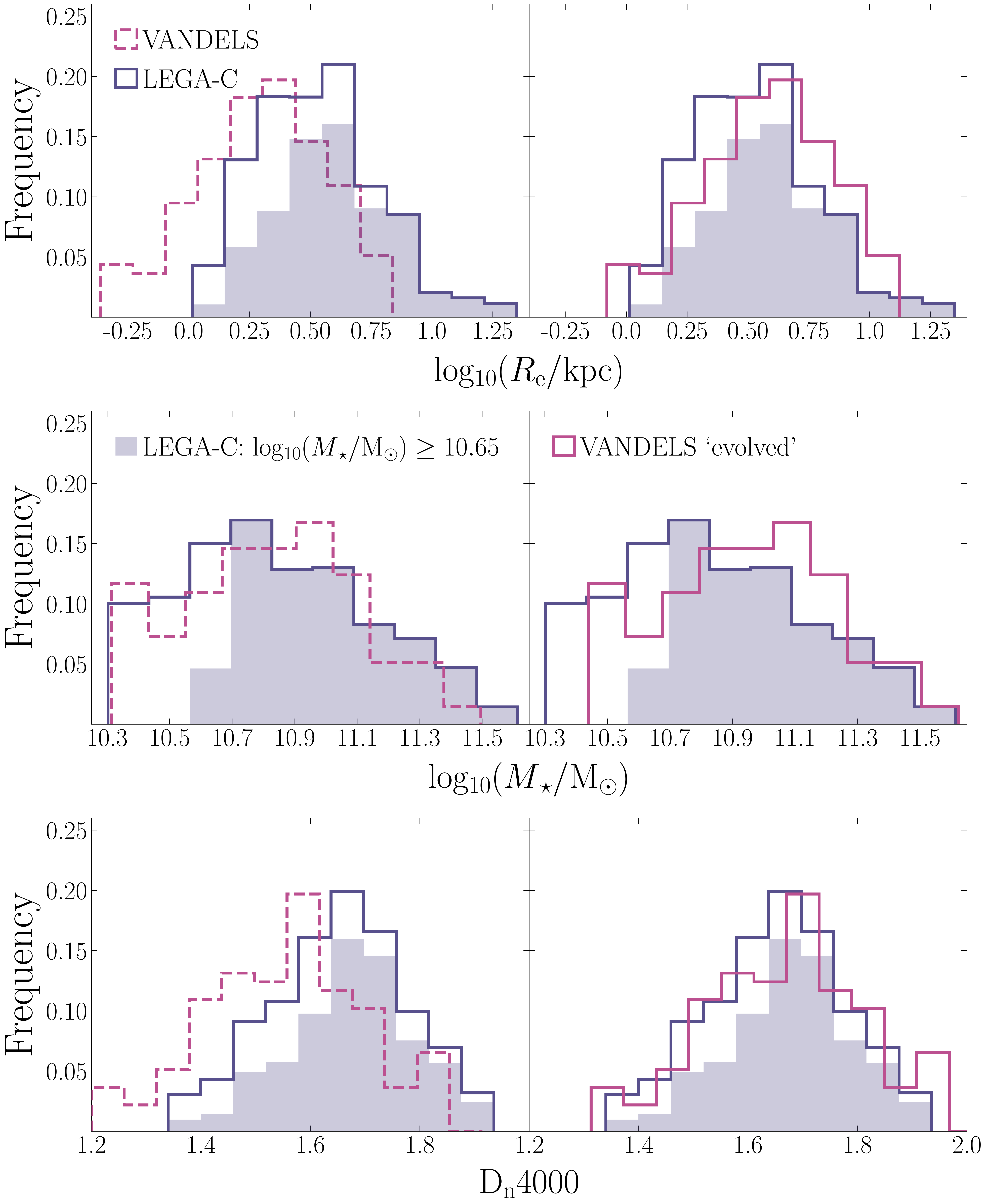}
   \caption{The distributions of size, stellar mass and D\textsubscript{n}4000 for the VANDELS (pink) and LEGA-C (indigo) galaxy samples. The left-hand panels show the observed distributions. In all panels, the grey shaded region corresponds to the subset of the LEGA-C sample with stellar masses $\mathrm{log_{10}}(M_{\star}/\mathrm{M_{\odot})} \geq 10.65$. This subset has the same number density as the full VANDELS sample with $\mathrm{log_{10}}(M_{\star}/\mathrm{M_{\odot})} \geq 10.3$. The right-hand panels show the observed VANDELS distributions shifted to the same median values as the corresponding LEGA-C distribution with $\mathrm{log_{10}}(M_{\star}/\mathrm{M_{\odot})} \geq 10.65$. The applied shifts are 0.28, 0.13 and 0.11 in $\log_{10}(R_{e}/{\rm kpc})$, $\mathrm{log_{10}}(M_{\star}/\mathrm{M_{\odot})}$ and D\textsubscript{n}4000, respectively. The histograms have been normalised by the total number of objects in each sample.}  
    \label{fig:evol_histograms}
\end{figure}

\subsubsection{\texorpdfstring{Modelling the evolution of size, stellar mass and D\textsubscript{n}4000}{Size, stellar mass and Dn4000 distributions}}\label{prob_legac_vandels}

In the left-hand panels of Fig. \ref{fig:evol_histograms}, we show size, mass and D\textsubscript{n}4000 histograms for the VANDELS (pink) and LEGA-C (indigo) samples. In order to explore whether the simplest set of assumptions possible can explain the trends observed in our data, we test a model wherein we calculate the shifts required to reconcile the two samples for each parameter, by finding the difference between the median of the VANDELS distribution and the median of LEGA-C galaxies with stellar masses $\mathrm{log_{10}}(M_{\star}/\mathrm{M_{\odot})}\geq 10.65$ (grey shading). In the right-hand panels of Fig. \ref{fig:evol_histograms}, we compare the original LEGA-C distributions to those of the shifted/evolved VANDELS sample, where the shifts applied in $\log_{10}(R_{e}/{\rm kpc})$, $\mathrm{log_{10}}(M_{\star}/\mathrm{M_{\odot})}$ and D\textsubscript{n}4000 are 0.28, 0.13 and 0.11, respectively.

The application of a two-sample Kolmogorov--Smirnov (KS) test \citep[][]{ks_test_scipy} confirms that the distributions shown in the right-hand panels of Fig. \ref{fig:evol_histograms} are statistically indistinguishable, with \textit{p}--values of 0.84, 0.56 and 0.70 for the size, mass and D\textsubscript{n}4000 distributions. In this case, a high \textit{p}--value suggests there is no statistical evidence that the two underlying distributions are significantly different. This confirms that, apart from a systematic shift, there is no significant change in the shape of the size, mass and D\textsubscript{n}4000 distributions between the LEGA-C and the VANDELS samples.

Moreover, we can test whether the implied change in the median D\textsubscript{n}4000 value of $\simeq 0.11$ is plausible, by comparing with the predictions shown in Fig.~\ref{fig:ssp_age_metallicity}. Within the range of plausible metallicities (i.e. $0.0 < [Z/H] <-0.3$), the tracks shown in Fig.~\ref{fig:ssp_age_metallicity} suggest that the change in D\textsubscript{n}4000 over 2 Gyr will indeed lie in the range \hbox{$0.1< \Delta {\rm D\textsubscript{n}4000}<0.2$}, as required. For massive quiescent galaxies at $z\sim0.8$, \cite{Tacchella2021} find a median metallicity of [Z/H] = -0.27 (on our Solar abundance scale), suggesting an expected D\textsubscript{n}4000 evolution of $\simeq0.1$ based on Fig.~\ref{fig:ssp_age_metallicity}. Although encouraging, we note that further spectrophotometric analysis of both datasets will be required to fully understand the evolution in age and metallicity between the two samples.

\begin{figure}
    \centering
    \includegraphics[width = \columnwidth]{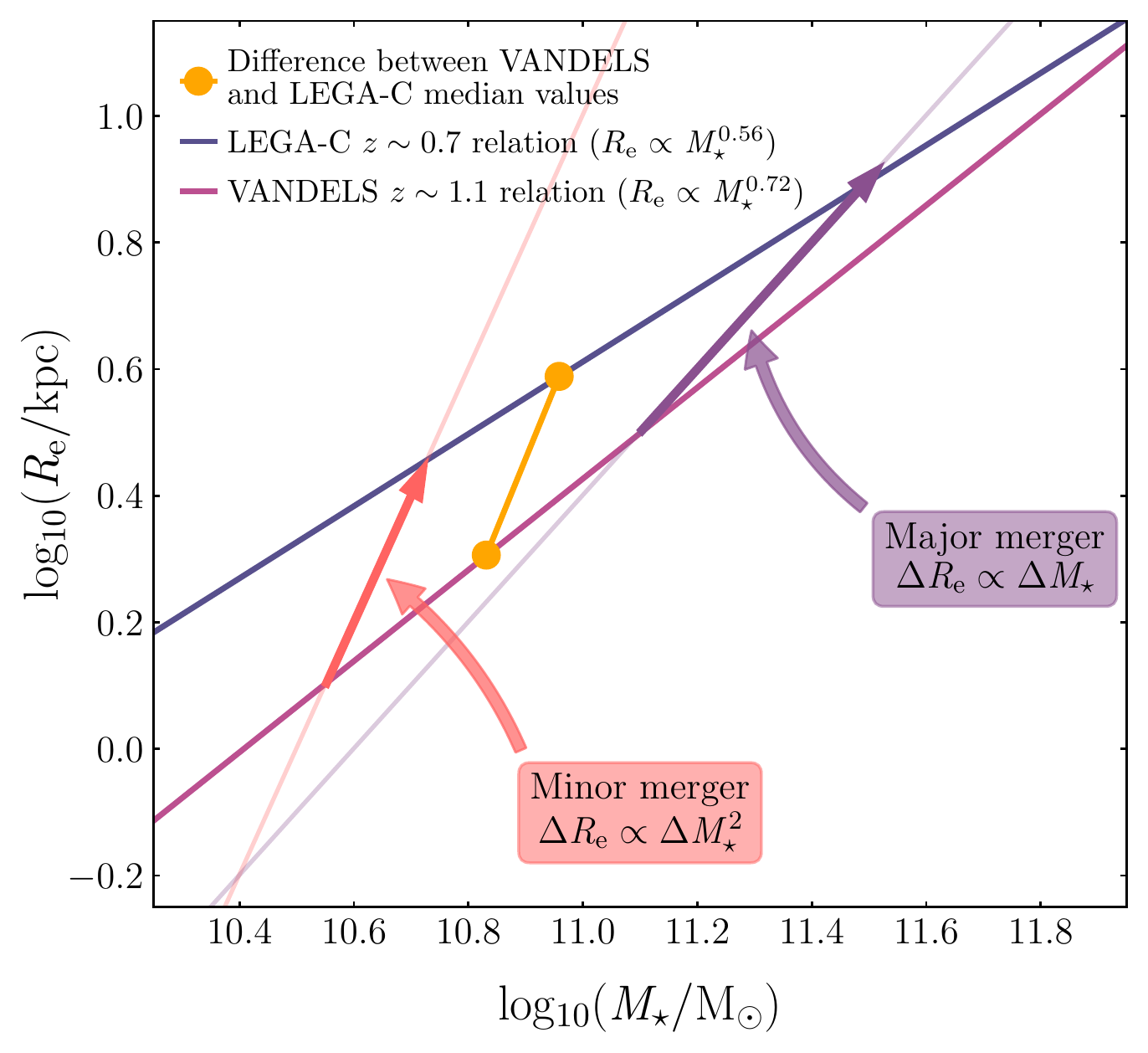}
    \caption{A schematic diagram showing the possible growth paths of the VANDELS galaxies as a result of major (purple) or minor (pink) mergers. The dark pink and indigo lines show the best-fitting mass-size relations derived in Section 4.2 for VANDELS and LEGA-C respectively. The joined orange circles show the median size and mass values for VANDELS galaxies with log$_{10}(M_{\star}/\mathrm{M_{\odot}}) > 10.3$, and for LEGA-C galaxies with log$_{10}(M_{\star}/\mathrm{M_{\odot}}) > 10.65$ (see Section \ref{toy_model_number_densities}). It can be seen that the median differences we observe between VANDELS and LEGA-C suggest that the size evolution of quiescent galaxies from $z\sim 1.1$ to $z\sim 0.7$ is dominated by minor mergers.}\label{fig:merger_scenarios}
\end{figure}

\subsubsection{Interpreting size growth via minor mergers}\label{minor_mergers}

It has long been accepted that the most likely explanation for the observed size growth of the quiescent galaxy population is via dissipationless, so-called `dry', mergers \citep[e.g.][]{trujillo_minor_mergers_2011,van_dokkum_minor_mergers_obs,cimatti_dry_merger,ross_size_2013}. The dry merging is usually classified as either `major', with a typical mass ratio of 1:3, resulting in size growth proportional to the accreted mass, or `minor', with a typical mass ratio of 1:10, resulting in size growth proportional to the square of the accreted mass \citep{cimatti_dry_merger}.

The potential growth of the VANDELS  galaxies via major and minor mergers is illustrated in the schematic diagram shown in \hbox{Fig. \ref{fig:merger_scenarios}}. The VANDELS and LEGA-C mass-size relations derived in \hbox{Section \ref{size-mass-relations}} are again shown, along with representative vectors showing the directions galaxies are expected to move under the minor-merger and major-merger scenarios. The two joined orange points show the median values calculated in Section \ref{prob_legac_vandels} for VANDELS galaxies, and for LEGA-C galaxies with stellar masses $\mathrm{log_{10}}(M_{\star}/\mathrm{M_{\odot})}\geq 10.65$. For context, these median values are also shown overplotted on the whole VANDELS and LEGA-C samples on the mass-size plane in Fig. \ref{fig:toy_model_evolved_vandels}.

Fig. \ref{fig:merger_scenarios} immediately suggests that the observed evolution between VANDELS and LEGA-C is more consistent with minor mergers. As discussed in the previous section, the observed growth between the VANDELS and LEGA-C samples in terms of stellar mass and size are 0.13~dex and 0.28~dex (factors of 1.35 and 1.91), respectively. This is clearly consistent with a minor-merger scenario (i.e. $\Delta R_{e}\propto \Delta M_{\star}^2)$. If we assume that minor mergers have a typical mass ratio of 1:10, the observed growth between VANDELS and LEGA-C is consistent with a series of three minor mergers within a time-span of $\simeq 2$ Gyr.

It is clear from simulations that typical galaxies in the mass range under discussion here are unlikely to experience a major merger over the redshift interval $0.7<z<1.1$ \citep[e.g.][]{hopkins_mergers_simulations,minor_mergers_sim_johannson}. This means that our results are consistent with the expectation that minor mergers are the dominant process driving quiescent galaxy growth at this epoch \citep[e.g.][]{ownsworth_mergers_2014,buitrago_2017}.

\subsubsection{Evolving the VANDELS sample down to $z\sim0.7$}\label{toy_model_evolution}

\begin{figure*}
    \centering
  \includegraphics[width = 0.95\textwidth]{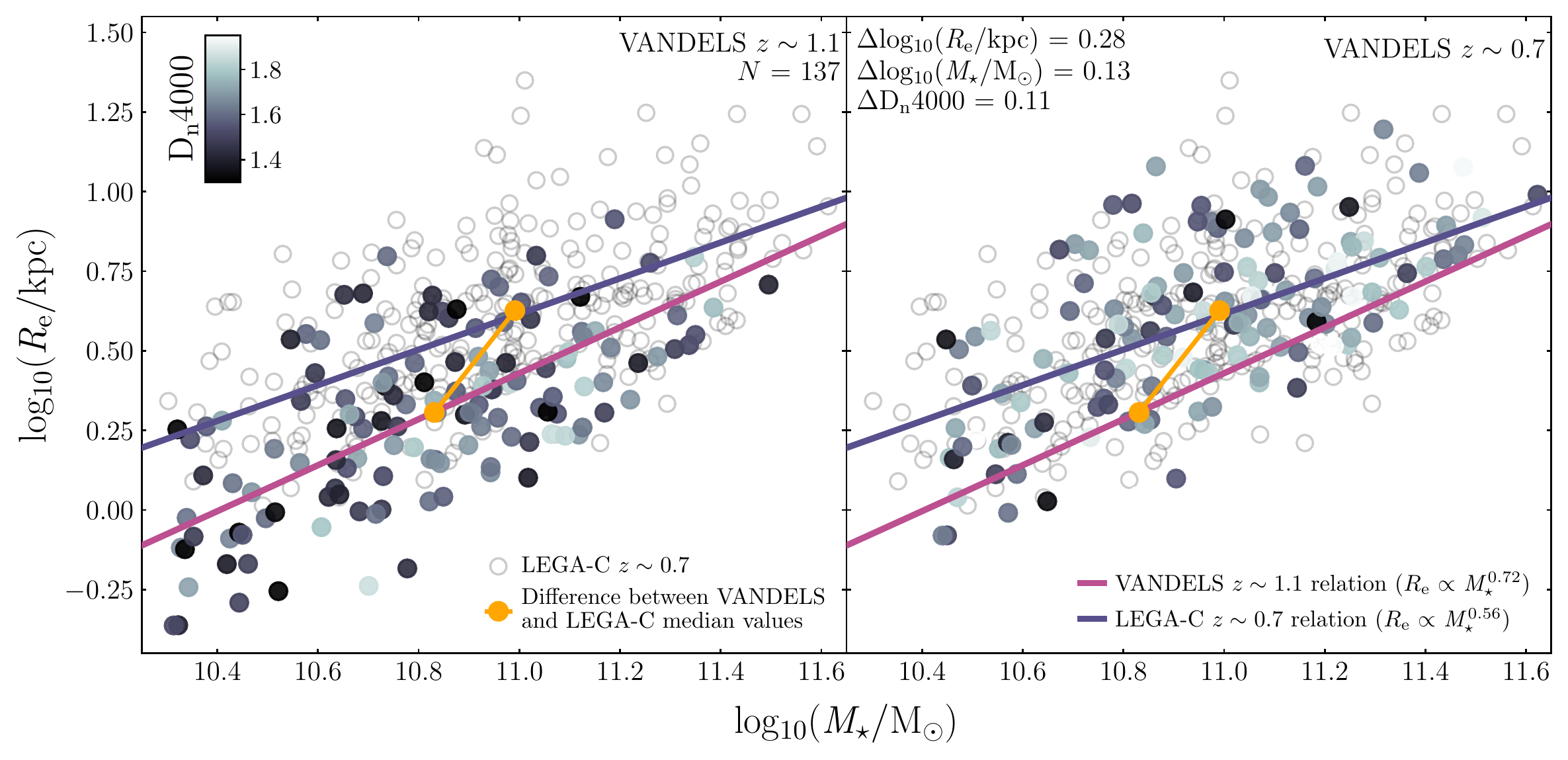}
\caption{Evolving the VANDELS galaxies down to $z\sim0.7$ on the mass-size plane. The left-hand panel shows the VANDELS galaxies (coloured by D\textsubscript{n}4000) overplotted on the LEGA-C sample (grey circles), both at their respective redshifts of $z \sim 1.1$ and $z \sim 0.7$. The right-hand panel shows the VANDELS galaxies shifted in stellar mass, size and D\textsubscript{n}4000 by the values shown in the top left-hand corner, derived in Section \ref{prob_legac_vandels}. In both panels, we also plot the best-fitting mass-size relations determined for the VANDELS and LEGA-C samples in Section 4.2. It can be seen that the mass-size distribution of the shifted VANDELS sample in the right-hand panel is indistinguishable from that of the LEGA-C sample at $\mathrm{log_{10}(\mathit{M_{\star}}/M_{\odot})}>10.65$ (see Section \ref{toy_model_evolution} for discussion).}
\label{fig:toy_model_evolved_vandels}
\end{figure*}

The expected future evolution of the VANDELS galaxies on the mass-size plane following our toy-model prescription is illustrated in \hbox{Fig. \ref{fig:toy_model_evolved_vandels}}. The left-hand panel shows the VANDELS sample at $z\sim 1.1$, along with the best-fitting mass-size relation determined in \hbox{Section \ref{size-mass-relations}}.  In the right-hand panel, we show the location of the VANDELS galaxies after they have been "evolved" down to $z\sim0.7$, using the average offsets from Section \ref{prob_legac_vandels}, as indicated in the top-left corner of the panel. The best-fitting mass-size relation for the \hbox{LEGA-C} sample at $z\sim 0.7$ is also shown. Individual LEGA-C galaxies are shown with grey open circles in both panels.

It can immediately be seen from the right-hand panel of Fig.~\ref{fig:toy_model_evolved_vandels} that the two-dimensional distribution of the evolved VANDELS sample is well matched to that of LEGA-C (confirmed by a 2D KS test, \citealt{peacock_2dKStest}; $p=0.96$). As discussed above, the bottom-right panel of \hbox{Fig. \ref{fig:evol_histograms}} demonstrates the two samples are also well matched in terms of their D\textsubscript{n}4000 distributions, and that the required D\textsubscript{n}4000 shift of 0.11 is consistent with 2 Gyr of passive ageing. Taken together, this suggests that our toy model, linking the two populations via a combination of passive evolution and minor mergers, is entirely plausible.

\section{Conclusions}\label{conclusions}
In this paper we have explored the relationships between stellar mass, size and D\textsubscript{n}4000 for samples of quiescent galaxies at $1.0 < z < 1.3$ and $0.6 < z < 0.8$, by utilising high-quality spectroscopic data from the VANDELS and LEGA-C surveys. Our main conclusions can be summarised as follows:

\begin{enumerate}
\setlength\itemsep{0.5em}
\item{In Section \ref{mass-d4000_relation}, we report a positive correlation between D\textsubscript{n}4000 and stellar mass in both the LEGA-C \hbox{($z\sim 0.7$)} and VANDELS \hbox{($z\sim1.1$)} samples, with a magnitude of $\simeq0.1$ across a $\simeq1$ dex interval in stellar mass. Within the mass and redshift ranges spanned by the two samples, we expect little or no correlation between stellar mass and metallicity \citep[e.g.][]{beverage_metallicity_not_age, Borghi_met_downsizing_30_06_21}. Consequently, we interpret this relationship between D\textsubscript{n}4000 and stellar mass as being primarily driven by a correlation between stellar mass and stellar-population age (downsizing).}
  
\item{In Section \ref{size-mass-relations}, we report a new mass-size relation for the VANDELS sample at $z \sim 1.1$, and confirm previous determinations of the mass-size relation for the LEGA-C sample at $z\sim 0.7$, with best-fitting slopes of \hbox{$\alpha$ = $0.72\pm0.06$} and \hbox{$\alpha = 0.56\pm0.04$}, respectively. Our results provide tentative evidence for a flattening in the slope of the mass-size relation towards lower redshift, although the level of flattening ($\Delta \alpha=-0.16\pm0.07$) is only significant at the $\simeq 2\sigma$ level.}
  
\item{In Section \ref{size_dn4000_results}, we find that, for the LEGA-C sample at fixed stellar mass, galaxies below the mass-size relation display larger D\textsubscript{n}4000 values than galaxies above the relation. This is in good agreement with the previous analysis of LEGA-C by \cite{wu_size_paper}. A similar trend is not clearly seen within the VANDELS sample at $z\sim 1.1$, although the smaller sample size and lower SNR of the individual VANDELS spectra would make detecting such a trend unlikely.}
  
\item{Our analysis suggests that the magnitude of the trend between D\textsubscript{n}4000 and size observed within the LEGA-C sample cannot be fully explained by a relationship between size and metallicity, meaning that a size-age relation must also be present in the data, with a magnitude of $\lesssim500$ Myr (see Section \ref{ssp_metallicity_age}).}
  
\item{We find that a simple toy model, based on a combination of passive evolution and minor mergers, can explain the observed evolution in stellar mass, size and D\textsubscript{n}4000 between the VANDELS and LEGA-C samples. This scenario, assuming each VANDELS galaxy experiences $\simeq 2$ Gyr of passive evolution, at a constant metallicity, together with a series of $N\simeq 3$ minor mergers, is sufficient to reproduce the distribution of the LEGA-C sample on the mass-size plane (see Section \ref{toy_model_section}).}

\end{enumerate}

\noindent In the near future, there are excellent prospects for improving our understanding of the evolution of size, mass, age and metallicity 
within the quiescent galaxy population at $z\geq 1$. In addition to the unparalleled near-IR imaging and spectroscopic data promised by forthcoming \textit{James Webb Space Telescope} programmes \citep[e.g. PRIMER;][]{primer}, large-scale optical-near-IR spectroscopic surveys, such as MOONRISE \citep[][]{moons}, will enable detailed studies of the quiescent galaxy population out to the highest redshifts.

\section*{Acknowledgements}
M. L. Hamadouche acknowledges the support of the UK Science and Technology Facilities Council. A. C. Carnall acknowledges the support of the Leverhulme Trust. FB acknowledges support from grants PID2020-116188GA-I00 and 
PID2019-107427GB-C32 from The Spanish Ministry of Science and Innovation. MM acknowledges support from MIUR, PRIN 2017 (grant 20179ZF5KS) and grants ASI n.I/023/12/0 and ASI n.2018-23-HH.0. Based on observations made with ESO Telescopes at the La Silla or Paranal Observatories under programme ID(s) 194.A-2003(E-Q) (The VANDELS ESO Public Spectroscopic Survey) and IDs 194-A.2005 and 1100.A-0949 (The LEGA-C ESO Public Spectroscopic Survey). Based
on data products from observations made with ESO Telescopes at the La
Silla Paranal Observatory under ESO programme ID 179.A-2005 and on
data products produced by TERAPIX and the Cambridge Astronomy Survey
Unit on behalf of the UltraVISTA consortium. This research made use of Astropy, a community-developed core Python package for Astronomy \citep{astropy:2013,astropy:2018}.

\section*{Data Availability}

The VANDELS survey is a European Southern Observatory Public Spectroscopic Survey. The full spectroscopic dataset, together with the complementary photometric information and derived quantities are available from \url{http://vandels.inaf.it}, as well as from the ESO archive \url{https://www.eso.org/qi/}. The \hbox{LEGA-C} data used in this paper are also publicly available from the ESO Science Archive. The LEGA-C spectra and catalogues are also available at \url{https://users.ugent.be/~avdrwel/research.html\#legac}. 
 
\color{black}
\bibliographystyle{mnras}
\bibliography{hamadouche2022}

\bsp
\label{lastpage}
\end{document}